\documentclass[prl,reprint,aps,superscriptaddress,longbibliography,floatfix]{revtex4-2}

\usepackage{graphicx}
\usepackage{bm}
\usepackage{bbm}
\usepackage{amsmath}
\usepackage{amssymb}
\usepackage{booktabs}
\usepackage{multirow}
\usepackage{enumerate}
\usepackage{mhchem}
\usepackage{upgreek}
\usepackage[dvipsnames]{xcolor}
\usepackage{comment}
\usepackage{ulem}
\usepackage{bbding}

\usepackage[hidelinks]{hyperref}
\hypersetup{colorlinks=true, linkcolor=blue, citecolor=red, urlcolor=blue}

\newcommand{\lp}{\left}
\newcommand{\rp}{\right}
\newcommand{\Abs}[1]{{\lp|{#1}\rp|}}

\newcommand{\ket}[1]{|{#1}\rangle}
\newcommand{\Ket}[1]{\lp|{#1}\rp\rangle}
\newcommand{\bra}[1]{\langle{#1}|}

\newcommand{\braoperket}[3]{\langle{#1}|{#2}|{#3}\rangle}
\newcommand{\Braoperket}[3]{\lp\langle{#1}\middle|{#2}\middle|{#3}\rp\rangle}
\newcommand{\up}{{\uparrow}}
\newcommand{\down}{{\downarrow}}

\newcommand{\0}{{\bm 0}}

\newcommand{\kk}{{\bm k}}
\newcommand{\pp}{{\bm p}}
\newcommand{\LL}{{\bm L}}
\newcommand{\RR}{{\bm R}}
\newcommand{\SSS}{{\bm S}}
\newcommand{\mcH}{{\mathcal{H}}}
\newcommand{\mcK}{{\mathcal{K}}}

\newcommand{\mcR}{{\mathcal{R}}}
\newcommand{\bbZ}{{\mathbbm{Z}}}

\begin{document}

\title{Symmetry-protected cubic-touching topological surface bands \\with tunable singularities}

\author{Jingtian Shi}
\affiliation{Materials Science Division, Argonne National Laboratory, Lemont, Illinois 60439, USA}

\author{Taylor L. Hughes}
\affiliation{Department of Physics, University of Illinois Urbana-Champaign, Urbana, Illinois 61801, USA}

\author{Ivar Martin}
\affiliation{Materials Science Division, Argonne National Laboratory, Lemont, Illinois 60439, USA}

\date{\today}

\begin{abstract}

We propose a class of topological surface bands in three-dimensional topological insulators that have symmetry-protected cubic-order band touching. Within the symmetry constraint, the band dispersion can continuously vary between cubic dispersion, moat band and multi-mini-valley structure with van Hove singularities by adjusting particle-hole asymmetry and anisotropy. Thus, there is a family of tunable density-of-state singularities ranging from power-law to logarithmic divergences. This offers a versatile platform for engineering strongly correlated phases of matter in topological surface states. We provide an example realization in a prism-lattice tight-binding model of angular-momentum-3/2 electrons.

\end{abstract}

\maketitle

\paragraph{Introduction.}

Topological crystalline insulators (TCIs) \cite{fu2011topological, ando2015topological} are insulators with in-gap edge or surface bands protected by crystal symmetries. In three dimensions (3D), the wide range of possible crystal symmetries allows for the protection of a variety of surface state dispersions. Indeed, previously identified symmetry-protected surface band types include Dirac cones \cite{hsieh2012topological, liu2013two, tanaka2013two}, quadratic band touchings \cite{fu2011topological}, nonsymmorphic-symmetry-protected hourglass fermions \cite{wang2016hourglass}, and fourfold Dirac fermions \cite{wieder2018wallpaper, zhou2021glide}. In light of the explosion of interest in strongly-correlated flat band systems \cite{sun2011nearly, bergholtz2013topological, peotta2015superfluidity, leykam2018artificial, kang2020dirac, yin2022topological, bistritzer2011moire, cao2018unconventional, cao2018correlated, chen2019evidence, tian2023evidence, cai2023signatures, lu2024fractional}, higher order band touchings having power-law-divergent density of states (DOS) (for $E_\kk \propto |\kk|^n$ in two spatial dimensions, DOS $\sim E^{-(n-2)/n}$, whereas saddle points generally have logarithmically-divergent DOS \cite{vanhove1953occurrence, yuan2019magic}) are of great interest as they provide platforms to engineer novel phases of matter in strongly correlated topological surface states.  

In general, for topological surface states that form as isolated two-band touching points, the maximum order of a band touching that can be protected by exact (i.e., not approximate/emergent) symmetries is cubic \cite{fang2012multiweyl}. Such cubic crossings have been predicted in surface quasi-particle bands of 3D topological superconductors \cite{fang2015tridirac} where the order of the crossing is protected by 3-fold rotational ($C_3$) and time-reversal ($T$) symmetries, in combination with the Bogoliubov-de Gennes (BdG) particle-hole ($P$) ``symmetry." Notably, the resulting cubic dispersion relation gives rise to a highly-divergent DOS $\sim E^{-1/3}$ that would amplify strong-correlation effects on the surface, but such correlations may be difficult to establish in superconducting quasiparticles. Here, since we are interested in establishing platforms for strongly-correlated 2D surface phases, we instead focus on the electronic insulator counterparts to superconductors that harbor cubic surface band structure.

\begin{figure}
	\centering
	\includegraphics[width=\columnwidth]{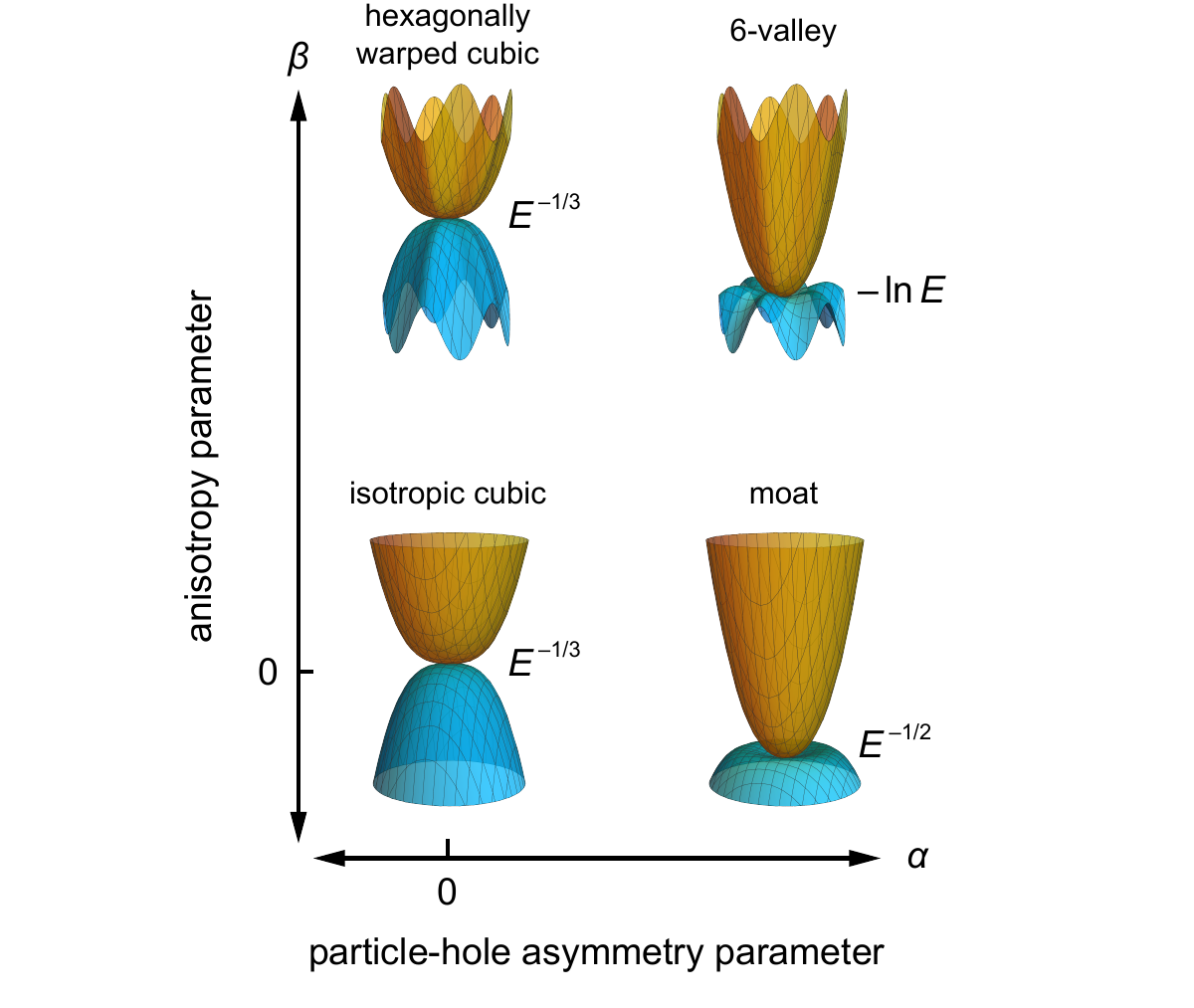}
	\caption{Schematics of various types of cubic-touching surface bands tuned by two phenomenological factors $\alpha$, $\beta$: $E_\kk \propto \alpha\kk^2 \pm \Abs{k_+^3 + \beta k_-^3}$, where $k_\pm = k_x\pm ik_y$. Order of DOS singularity is marked for each case.}
	\label{fig:intro}
\end{figure}

Unlike BdG Hamiltonians for superconducting quasiparticles, in the context of electronic insulators, particle-hole symmetry is not enforced. This gives extra versatility to the design and tunability of symmetry-protected topological surface band dispersions. To this end, here we identify a variety of surface band dispersion types whose band touchings have cubic order, i.e., $E_\kk^{\rm upper} - E_\kk^{\rm lower} \propto |\kk|^3$ (possibly up to some anisotropy factor $f(\theta_\kk)$). As the summary Fig. \ref{fig:intro} shows, the resulting surface state types include isotropic and anisotropic cubic dispersions, both with DOS singularity $\sim E^{-1/3}$. These dispersions can be tuned via a symmetry-allowed, flavor-independent quadratic term $\alpha\kk^2$ which is not available in the superconductor quasiparticle context. As an example of the effect of such a term, starting from the isotropic cubic case, for $\alpha >0\, (\alpha<0)$ the upper (lower) band develops ring minimum (maximum), i.e., moat dispersion \cite{berg2012electronic, sedrakyan2014absence}, enhancing the DOS singularity to $\sim E^{-1/2}.$ If instead we start with the anisotropic, hexagonally warped cubic, the ring minimum (maximum) is further hexagonally warped to form six mini-valleys around  six discrete minima (maxima) and with six saddle points between them. 
The most general band structure of this type that requires the lowest symmetry is the ``6-valley'' type. This dispersion can be protected by 3-fold $C_3$  (6-fold $C_6$) rotational symmetries in combination with time-reversal symmetry $T$ (reflection symmetry about a 
crystallographic mirror $M$ perpendicular to the surface). Other types of dispersions can be achieved by tuning parameters to regions having less-generic symmetries such as particle-hole symmetry $P$, chiral symmetry $S = TP,$ and/or the (approximate) continuous rotational symmetry generated by $J_z$, as summarized in Table \ref{table:sym_summary}. For certain symmetries the cubic-touching can split into three Dirac cones while still remaining gapless. In that case, there are also saddle points between the Dirac cones, which in some cases can fuse into higher-order (``monkey") saddles \cite{shtyk2017electrons}, leading to higher van Hove singularities (VHSs). We will show that all of these cases can be realized in the context of a single model with electrons having effective angular momentum $J_z \equiv 3/2$ (mod 3), a property that turns out to be essential to support cubic touching dispersions. 

\begin{table}
\caption{Types of topological surface band dispersions and their protecting symmetries represented by generator lists of symmetry groups, assuming that the $\kk = \0$ space for the surface bands carries the $|j_z| = 3/2$ representation of the symmetry group with $T^2 = -1$ and $P^2 = 1$. Only ``3 Dirac cones'' lacks cubic touching.}
\begin{tabular}{|p{0.34\columnwidth}|p{0.62\columnwidth}|}
	\hline
	dispersion type & protecting symmetries  \\\hline
    \raggedright isotropic cubic & $\{J_z, S\}$ or $\{J_z, PM\}$ \\\hline
    \raggedright moat & $\{J_z, T\}$ or $\{J_z, M\}$ \\\hline
    \raggedright hexagonally warped cubic & $\{C_3, SC_2\}$, $\{C_3, T, P\}$, $\{C_3, T, PM\}$, $\{C_3, P, M\}$, or $\{C_6, PM\}$ \\\hline
	\raggedright 6-valley & $\{C_3, T\}$ or $\{C_6, M\}$ \\\hline\hline
	\raggedright 3 Dirac cones & $\{C_3, S\}$ or $\{C_3, TC_2\}$ \\\hline
\end{tabular}
\label{table:sym_summary}
\end{table}

\paragraph{General symmetry analysis of surface bands.}

In order to obtain symmetry constraints for higher  order band touching points, let us consider
 general $2\times2$ $\kk\cdot\pp$ Hamiltonian for a topological surface state:
\begin{equation}
\begin{aligned}
    \mcH(\kk) &= \sum_{n_+, n_-} k_+^{n_+} k_-^{n_-} \\
    &\times \bigl( h_{n_+,n_-}^0\sigma_0 + h_{n_+,n_-}^z\sigma_z + h_{n_+,n_-}^+\sigma_+ \bigr) + H.c.,
    \label{eq:k.p_general}
\end{aligned}
\end{equation}
centered around the $\Gamma$-point. Here $k_\pm = k_x \pm ik_y$, $\sigma_0$ is the $2\times2$ identity matrix, $\sigma_{\pm} = \sigma_x \pm i\sigma_y,$ and $h_{n_+,n_-}^{0,z,\pm}$ are complex coefficients. The order of the band touching is determined by the minimal value of $n_+ + n_-$ in the $\sigma_z$ and $\sigma_+$ terms. Note that at this stage $\sigma_a$ may correspond to any set of two degrees of freedom, not necessarily physical spin, and the symmetry representations will determine the interpretation of $\sigma_a.$ We start from the case where continuous rotation $J_z$, mirror $M$, time-reversal $T$, and (fine-tuned) particle-hole $P$ symmetries are all present, so that the Hamiltonian is constrained by
\begin{subequations}
\label{eq:k.p_sym_general}
\begin{align}
    T\mcH(\kk)T^{-1} &= \mcH(-\kk), \\
    P\mcH(\kk)P^{-1} &= -\mcH(-\kk), \\
    M\mcH(k_x, k_y)M^{-1} &= \mcH(k_x, -k_y), \\
    e^{-i\theta J_z}\mcH(\kk)e^{i\theta J_z} &= \mcH(\mcR_\theta\kk),
\end{align}
\end{subequations}
where $\mcR_\theta$ rotates a 2D real vector counterclockwise by angle $\theta$. To fix the matrix forms of the symmetries, we assume that the Hilbert space at the $\Gamma$-point carries a fermionic representation having angular momentum quantum number $|j_z| =n/2,$ where $n$ is an odd integer \cite{supplemental1}. One convenient choice of the matrix representation of the symmetries is
\begin{equation}
    T = i\sigma_y\mcK, \quad P = \sigma_x\mcK, \quad M = \sigma_x, \quad J_z = \frac{n}{2}\sigma_z,
    \label{eq:surface_syms_general}
\end{equation} where $\mcK$ is complex conjugation.

Let us now identify the surface state dispersions allowed by these symmetries.
In the most restrictive case, when all symmetries are enforced, we see that chiral symmetry $S = TP = \sigma_z$ forbids all $\sigma_0$ and $\sigma_z$ terms in Eq. (\ref{eq:k.p_general}), while the continuous rotation symmetry $e^{-i\theta J_z}$ further limits the allowed $\sigma_+$ terms to have momentum dependence satisfying $n_ - -n_+ = n.$ Thus, the lowest-order term allowed would give an isotropic, $n$th-order band dispersion, e.g., $n=3$ will yield an isotropic cubic band touching dispersion. Lifting particle-hole symmetry $P$ allows for some $\sigma_0$ and $\sigma_z$ 
terms. However, rotation symmetry will require $n_+ = n_-$ for any $\sigma_0$ or $\sigma_z$ terms. Moreover, the latter are forbidden by $T$ or $M$. Hence, the $n$th-order relative touching $E_\kk^+ - E_\kk^-,$ and isotropy, are not affected by simply lifting $P$. 

Next, consider reducing $J_z$ to discrete rotational symmetry $C_\nu = e^{-2\pi iJ_z/\nu} = e^{-i\pi n\sigma_z/\nu}.$ This allows new terms with values of $n_--n_+$ differing by multiples of $\nu$. When $\nu=2n$, this still forces $|n_+ - n_-| \ge n$ for the $\sigma_+$ terms, thus the $n$th-order band touching would be protected. Note, however, that now coexistence of $k_+^n\sigma_+$ and $k_-^n\sigma_+$ terms can lead to anisotropy in the $n$th-order dispersion, e.g., hexagonally warped cubic in the case of $n=3$. In crystals, the highest possible exact rotational symmetry is $C_\nu=C_6$, hence the highest order band touching protected by exact symmetries is cubic ($n=3$), which is realized for $J_z \equiv 3/2$ (mod 3). All other angular momenta having mod 3 remainder not equal to $3/2$ would stabilize only lower-order band touching points; all higher discrete or continuous rotational symmetries have to be approximate in the $\kk\cdot\pp$ limit, and realized only by fine tuning of model parameters. A full symmetry analysis for $|j_z| = 3/2$ is presented in the Supplemental Material (SM) \cite{supplemental1}, assuming at least $C_3$ symmetry, while the results are summarized in Table \ref{table:sym_summary}.

\begin{figure}
	\centering
	\includegraphics[width=0.48\textwidth]{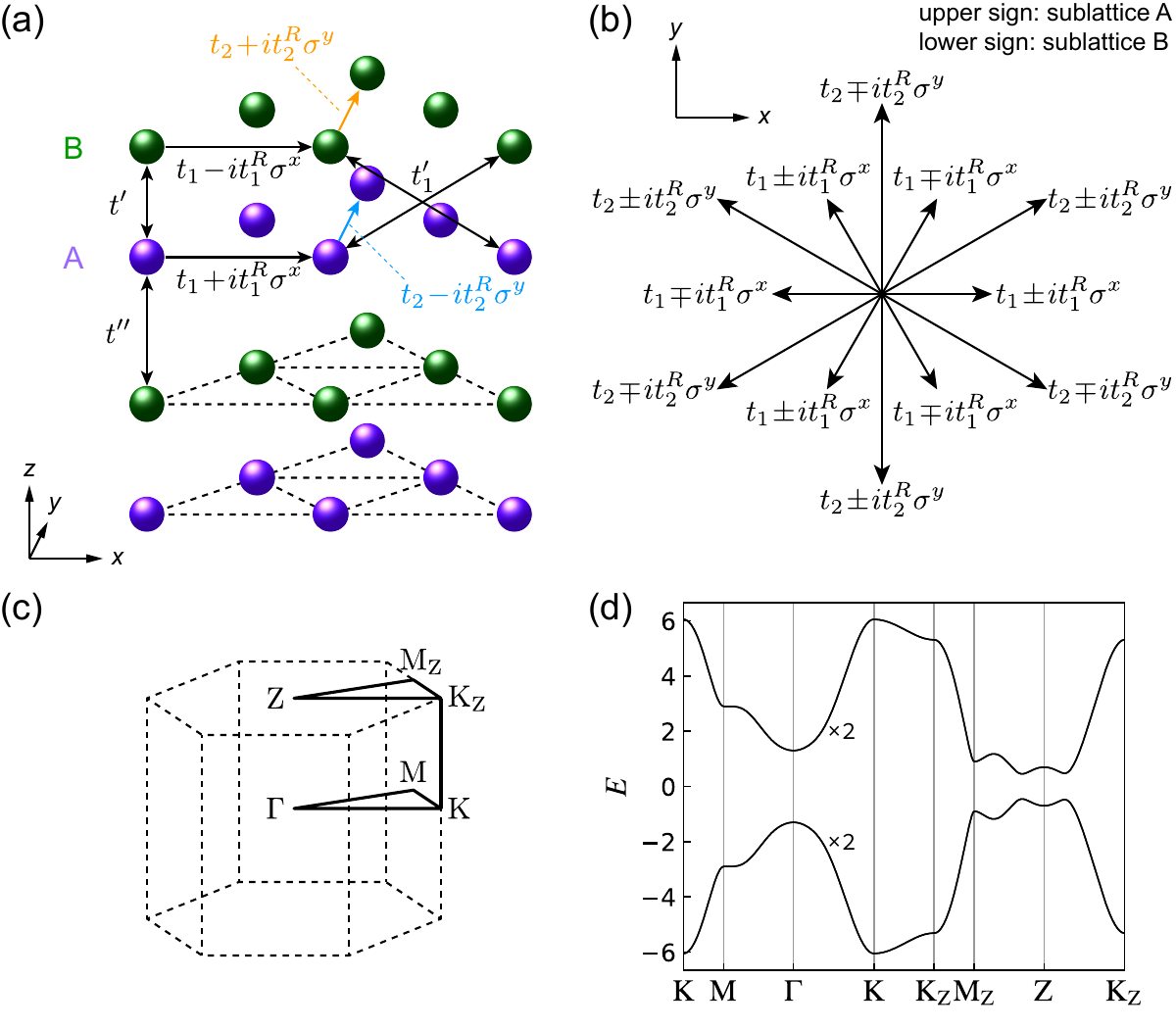}
	\caption{(a) An illustration of the triangular prism lattice tight-binding model in Eq. (\ref{eq:bulkH}), with inter-site tunneling matrices represented with arrows. The coordinate frame is shown on the bottom left. (b) All tunneling matrices within the $xy$-plane. (c) The 3D Brillouin zone (BZ) of the system. (d) A typical 3D band structure of the model along the solid lines shown in (c), in which both bands are doubly degenerate.}
	\label{fig:model}
\end{figure}

\smallskip
\paragraph{Tight-binding model.}
To provide an example of a system with tunable symmetry-protected cubic touching in topological surface bands, we construct a prism lattice with two inequivalent sites $A$ and $B$ along the $z$-axis in each unit cell, as illustrated in Fig. \ref{fig:model}(a). We place a $j_z = \pm3/2$ electron on each site, assuming that the energies of $j_z = \pm1/2$ are shifted away so that they do not mix with the $j_z = \pm3/2$ orbitals. Note that the lattice geometry has sublattice-preserving symmetries $C_6$ and $M$ ($y\to -y$), as well as the sublattice-exchanging 3D spatial inversion symmetry $I$. Combining them with the spin-3/2 representation of angular momentum, we find the physical symmetries of the bulk tight-binding model:
\begin{equation}
    T = i\sigma^y\mcK, \quad C_3 = \mathbb{I}, \quad C_2 = \sigma^z, \quad M = \sigma^x, \quad I = \tau^x,
    \label{eq:bulk_syms}
\end{equation}
(up to arbitrary phase factors), where $\mathbb{I}$ is the identity matrix, $\tau$ and $\sigma$ denote sublattice and spin Pauli matrices respectively, and we use superscripts to distinguish from the subscripted Pauli matrices in the surface band basis we used in the previous section. 

To parameterize a tight-binding model we identify  Hamiltonian terms allowed by the symmetries listed in Eq. (\ref{eq:bulk_syms}). A detailed analysis is presented in the SM \cite{supplemental1}. We identify the following allowed near-neighbor terms in the momentum-space Bloch Hamiltonian, whose corresponding real-space hopping terms are illustrated in Figs. \ref{fig:model} (a), (b):
\begin{equation}
	H(\kk) = H_P(\kk) + H_0(\kk) = H_P(\kk) + \sum_{i=0}^3 h_i(\kk) \Gamma^i,
    \label{eq:bulkH}
\end{equation}
where
\begin{equation}
	H_P(\kk) = t_1 c_1(k_x, k_y) + t_2 c_2(k_x, k_y),
    \label{eq:HP}
\end{equation}
\begin{subequations}
\label{eq:h}
\begin{align}
	h_0(\kk) &= t' + t_1'c_1(k_x, k_y) + t'' \cos k_z, \label{eq:h0} \\\addlinespace
	h_1(\kk) &= t_1^R s_1(k_x, k_y), \label{eq:h1} \\\addlinespace
	h_2(\kk) &= t_2^R s_2(k_x, k_y), \label{eq:h2} \\\addlinespace
	h_3(\kk) &= t''\sin k_z, \label{eq:h3}
\end{align}
\end{subequations}
\begin{equation}
	\Gamma^0\!=\!\tau^x, \;\; \Gamma^1\!=\!\tau^z\sigma^x, \;\; \Gamma^2\!=\!\tau^z\sigma^y, \;\; \Gamma^3\!=\!\tau^y,
    \label{eq:Gamma_defs}
\end{equation}
and
\begin{equation}
    c_j(\kk_\perp) = 2\sum_{i=0}^2 \cos (\kk_\perp\!\cdot\!\RR_i^j), \;\;
    s_j(\kk_\perp) = 2\sum_{i=0}^2 \sin (\kk_\perp\!\cdot\!\RR_i^j),
    \label{eq:c_s}
\end{equation}
for $j = 1, 2$ and $\kk_\perp = (k_x, k_y)$; $\RR_i^1 = (\cos\frac{2\pi i}{3},\, \sin\frac{2\pi i}{3})$ and $\RR_i^2 = \sqrt{3}(\sin\frac{2\pi i}{3},\, -\!\cos\frac{2\pi i}{3})$. For convenience, we have assumed a unit lattice constant in the triangular layers and along the $z$-axis. The direct in-plane hopping amplitudes $t_1$ and $t_2$, the in-plane spin-orbit coupling/Rashba hoppings $t_1^R$ and $t_2^R$, the direct interlayer hoppings $t'$ and $t''$, and  the skew hopping $t_1'$ (illustrated in Fig. \ref{fig:model}(a)) are real parameters of the model. We see that the Bloch Hamiltonian has the symmetries in Eq. (\ref{eq:bulk_syms}), and that $H_0(\kk)$ has an extra particle-hole symmetry $PH_0(\kk)P^{-1} = -H_0(-\kk)$ where $P = \tau^z\sigma^x\mcK$. Note that $\{P, T\} = \{P, C_2\} = [P, C_3] = [P, M] = 0$ from Eq. (\ref{eq:bulk_syms}), which is consistent with the algebra for the surface Hamiltonian in Eq. (\ref{eq:surface_syms_general}) (there, $C_3 = e^{-2i\pi J_z/3}$). Hence, $H_0$ has the symmetry group representation necessary for protection of cubic touching as discussed in the previous section. Along the high-symmetry lines shown in Fig. \ref{fig:model}(c), a typical bulk band structure of $H_0(\kk)$ is plotted in Fig. \ref{fig:model}(d) with the parameter choice $t_1^R = 1$, $t_2^R = 0.2$, $t' = 1.5$, $t'' = 1$, $t_1' = -0.2$, which belongs to the regime of our intended TCI phase as we will see below. The universal double degeneracy of the band structure is protected by the combined symmetry $K = IT = i\tau^x\sigma^y\mcK$, which is like a momentum-preserving Kramers symmetry. We also note that $H_P(\kk)$ violates $P$, and hence tunes the particle-hole asymmetry as discussed in the introduction.

\begin{figure*}
	\centering
	\includegraphics[width=\textwidth]{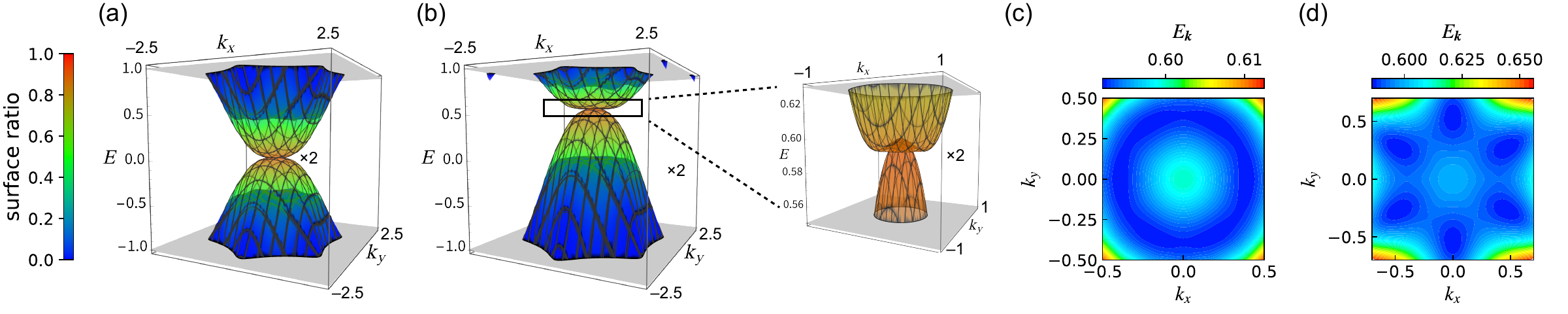}
	\caption{(a) The band structure of an $xy$-slab containing 50 unit cells in the $z$-direction, whose corresponding bulk Hamiltonian is $H_0(\kk)$ with $t' = 1.5$, $t'' = 1$, $t_1' = -0.2$, $t_1^R = 1,$ and $t_2^R = 0.2$. The color scale represents the surface vs. bulk character of the states where warmer colors toward red indicate surface states, and cooler colors toward blue represent states with a bulk character.  (b) The same as (a) but with the particle-hole-symmetry-breaking term $H_P(\kk)$ added with $t_1=0.1$ and $t_2=0$. The zoomed-in band touching structure is shown on the right. (c) The 2D constant energy contours of the first surface conduction band in (b), \textit{i.e.}, the first band above the touching point, within a small range centered at $\Gamma$. (d) The energy contours of the same band with the same parameters as in (c) except that $t_2^R = 0.15$.}
	\label{fig:surfacestates3Dplot}
\end{figure*}

\smallskip
\paragraph{Surface bands}
To illustrate the tunable surface state dispersions we compute the energy spectrum in a slab geometry. Fig. \ref{fig:surfacestates3Dplot}(a) shows the band structure of a slab that is periodic in the $xy$-plane and contains 50 unit cells along the (open boundary) $z$-direction. We use a bulk Hamiltonian $H_0(\kk)$ with the same parameters as the bulk band structure shown in Fig. \ref{fig:model}(d). We see two degenerate pairs of surface bands touching at the 2-dimensional (2D) BZ center ($\Gamma$), with one set coming from each surface, and we find that the surface bands have cubic dispersions near the $\Gamma$-point. Band structures using other interlayer hopping parameter choices can display different forms of surface bands. For example, in the SM \cite{supplemental1} we illustrate the presence of surface bands associated with DIII strong topology, 1-dimensional (1D) weak topology, and/or mirror Chern numbers.
Turning on the particle-hole-symmetry-breaking term $H_P$ introduces an overall quadratic bending to the cubic-touching structure, leading to moat-like (approximate) ring minimum in the surface band, as illustrated in Figs. \ref{fig:surfacestates3Dplot} (b), (c). Fig. \ref{fig:surfacestates3Dplot}(d) shows that decreasing $t_2^R$ makes the surface band dispersion less isotropic, and breaks the ring minimum into 6 mini-valleys and 6 saddle points. Thus, we see how tuning the parameters of our bulk tight-binding model takes us through the entries of Table \ref{table:sym_summary}.
The divergence in the surface DOS is also verified in SM \cite{supplemental1}.

To understand these results we can construct the $\kk\cdot\pp$ Hamiltonian for the surface states. We note that at the $\Gamma$-point the system is effectively two spin-degenerate copies of the Su-Schrieffer-Heeger (SSH) chain \cite{su1979solitons}, whose edge modes on the top (bottom) surface have weight on only sublattice B (A)\cite{asboth2016short}. We construct the $\kk\cdot\pp$ Hamiltonian $\mcH^{\rm surf}(\kk)$ starting from these degenerate boundary modes. First consider the case with $P$ symmetry. The skew-hopping term $t_1c_1(k_x, k_y)\tau^x$ (see Eq. (\ref{eq:h0})) is quadratic in $\kk$ in the $\kk\cdot\pp$ limit, but does not contribute to $\mcH^{\rm surf}(\kk)$ because it couples different sublattices, and hence, from the remark above, does not couple states on the same surface. The only $(k_x, k_y)$-dependent terms left, $h_1(\kk)\Gamma^1$ and $h_2(\kk)\Gamma^2$ (see Eqs. (\ref{eq:h1}) and (\ref{eq:h2})), are third-order in $\kk$, ensuring the cubic band touching. Given the forms of $\Gamma^i$ in Eq. (\ref{eq:Gamma_defs}), the $\kk\cdot\pp$ Hamiltonian directly inherits the spin structure of those two terms. Thus, to the leading order we find
\begin{equation}
    \mcH^{\rm surf}(\kk) = -\frac{t_1^R}{8} \lp( \frac{1+\gamma}{2} k_-^3 + \frac{1-\gamma}{2} k_+^3 \rp)\sigma_+ + H.c.,
    \label{eq:mcH0_prism}
\end{equation}
where $\gamma = 3\sqrt{3}\,t_2^R/t_1^R$. In comparison to Eq. (\ref{eq:k.p_general}), when $\gamma = \pm1$, $\mcH^{\rm surf}(\kk)$ has a perfectly isotropic cubic band touching, corresponding to the case where the long-wavelength emergent continuous rotational symmetry generated by $J_z$ discussed earlier is present. This explains the difference in isotropy of Figs. \ref{fig:surfacestates3Dplot} (c) and (d). Moreover, turning on the term $H_P(\kk)$ in the bulk simply introduces a $(k_x, k_y)$-dependent overall shift to the surface state band structure. We can project the bulk symmetry representation onto the surface states to arrive at the surface symmetry representation: $P=\sigma_x\mcK,$ $T=i\sigma_y\mcK$, $C_3=1$, $C_2=\sigma_z$, and $M=\sigma_x$, which are consistent with Eq. (\ref{eq:surface_syms_general}) up to phase factors (inversion symmetry interchanges the two surfaces, and hence does not have a local surface projection). This enables us to understand the properties of the surface Hamiltonian using the general principles of our symmetry protection arguments for cubic band touching to this system.

\paragraph{Summary and discussion}

We have identified a class of symmetry-protected cubic-touching topological surface bands in $C_3$-symmetric 3D topological insulators with various degrees of divergence in their DOS.
In an earlier study, non-symmetry-protected cubic dispersion in 1D edge channels of 2D topological insulators with prescribed semi-Dirac bulk band structures was predicted \cite{olmos2026cubic}.
In our work, the types of divergence range from cubic dispersion $\sim E^{-1/3}$, to moat-like dispersion $\sim E^{-1/2},$ to discrete mini-valleys with saddle-point singularities $\sim -\ln E$, which our model can tune between using parameters that control particle-hole asymmetry and anisotropy. The role of cubic touching is crucial to the DOS divergence, because it ensures that when a symmetry-allowed, perturbative quadratic kinetic energy $\sim \alpha\kk^2$ is turned on, the extrema of at least one band shift away from the touching point. This, in turn, leads to moat-like dispersions, or VHSs, depending on the anisotropy. In contrast, quadratic band-touching points do not generally develop divergent DOS under such perturbations. In addition, certain violations of exact symmetries can also preserve the DOS singularities by splitting the cubic-touching point into 3 Dirac cones, an effect analogous to trigonal warping effects on the chiral electrons in rhombohedral trilayer graphene \cite{koshino2009trigonal, zhang2010band, jung2013gapped}, in which VHSs are also present in the band dispersions. More details about this case are included in the full surface symmetry analysis in the SM \cite{supplemental1}. We point out, however, that symmetry analysis alone does not fully determine the type of surface band dispersion -- for example, there can be accidental degeneracies or extra symmetries at high-symmetry points that lead to the dispersion types listed in Table \ref{table:sym_summary}, even if the symmetry group of the full Hamiltonian does not belong to any combination of symmetries listed therein \cite{supplemental1}.

Beyond the symmetry analysis, we have provided a tight-binding model realization of  cubic-touching topological surface states in a prism lattice using $j_z = \pm 3/2$ orbitals. Moreover, we demonstrated that the surface states can be tuned through the various cubic-touching types in Table \ref{table:sym_summary}.
However, in order to keep the bulk system an insulator, and to keep the surface band DOS singularity inside the bulk gap, $|t_1|$ has to be small compared to $|t_1^R|$ in our model. Otherwise, a large energy overlap between the bulk valence and conduction bands will occur, making the system a semimetal. In real materials, the Rashba term is typically small compared to ordinary, spin-independent hopping.
However, it has been demonstrated that some tellurium compounds can be well-described by models where the spin-independent hopping is small compared to Rashba coupling\cite{liu2014weyl}.
Additionally, in the SM \cite{supplemental1} we provide (i) an example which demonstrates the DOS singularity while preserving the conventional physical hierarchy between $|t_i|$ and $|t_i^R|$ at the cost of introducing further-neighbor hoppings, and (ii) a route to suppress spin-independent hopping relative to the Rashba coupling by assembling the $j_z = \pm3/2$ states from $p$ and $d$ atomic orbitals.

Alternatively, cold atoms \cite{liu2023elementary} in optical and/or Raman lattices provide plausible routes to engineering of our simpler model in the parameter regime of interest. The elements we need: in-plane triangular optical lattices \cite{becker2010ultracold}, dimerized perpendicular hopping \cite{atala2013direct, leder2016realspace}, spin-3/2 atoms \cite{wu2003exact} with $j_z = \pm 1/2$ energies lifted away by quadratic Zeeman effect \cite{rodriguez2010mottinsulator}, spin-orbit coupling \cite{wu2016realization}, and laser-assisted complex hopping patterns \cite{aidelsburger2013realization, miyake2013realizing} have been demonstrated. The last piece, a Rashba coupling that is independently tunable via microwave driving and lattice shaking \cite{grusdt2017tunable} has been proposed, potentially enabling a reversal of the scale hierarchy between the Rashba term and the spin-independent hopping. Thus, cold-atomic gases may offer a platform where the surface state types can be tuned {\it{in-situ}.}

Being able to continuously vary between different types of singular surface-band dispersions provides a platform to access strongly-correlated topological surface states upon inclusion of electron-electron interactions. For example, at or near charge neutrality, the high-order band touching structure may favor excitonic insulator states \cite{nandkishore2010dynamical}; the quadratically bent cubic touching can lead to the coexistence  of an annular electronic Fermi pocket, with a smaller hole Fermi pocket inside; anisotropy can further induce a Lifshitz transition and a splitting of the annular Fermi pocket into 6 mini-pockets. These may precipitate the interplay and competition between unconventional superconductivity \cite{zhou2021superconductivity, ghazaryan2021unconventional, qin2023functional}, finite-momentum excitonic pairing \cite{wang2023excitonic}, and mini-valley-biased nematic orders \cite{huang2023spin, mayrhofer2026nematic}, among others. Exact precedents of these fermiologies do not exist to our knowledge, thus we leave a systematic study of many-body phases in our topological surface bands for future work.

{\it Acknowledgements:}
J.S and I.M. were supported by the Center for the Advancement of Topological Semimetals (CATS), an Energy Frontier Research Center funded by the U.S. Department of Energy (DOE) Office of Science (SC), Office of Basic Energy Sciences (BES), through the Ames National Laboratory under contract DE-AC02- 07CH11358. TLH thanks ARO MURI W911NF2020166 for support.
Part of analysis and discussion in this work is assisted by OpenAI Codex with GPT-6-Astra model.

\bibliography{bibliography_revised}

\clearpage

\appendix
\onecolumngrid
\renewcommand\theequation{S\arabic{equation}}
\renewcommand\thefigure{S\arabic{figure}}
\setcounter{equation}{0}
\setcounter{figure}{0}

\section*{Supplemental material for ``Symmetry-protected cubic-touching topological surface bands with tunable singularities''}

\subsection{Representation of symmetry group}

We identify 2D representations of the full symmetry group generated by $T$, $P$, $M$ and $J_z$, assuming that its spatial and time-reversal parts directly inherit from the representation of SU(2) group with angular momentum quantum number $J$ and the corresponding particle statistics. Since both $T$ and $M$ flip the $J_z$ eigenvalue (noting that $z$-axis is in the mirror plane of $M$, and that angular momentum is invariant under 3D spatial inversion), any pair $\{\ket{m}, \ket{-m}\}$ for nonzero projected angular momentum quantum number $m$ is a representation basis, under which $J_z = m\sigma_z$, $T = \sigma_x\mcK$ when $J$ is an integer (\textit{i.e.} in the bosonic case) and $T = i\sigma_y\mcK$ when $J$ is half of an odd-integer (\textit{i.e.} in the fermionic case), and $M$ can only be $\sigma_x$, $\sigma_y$ or any of their linear combination satisfying $M^2=1$. Since all allowed choices of $M$ are gauge-equivalent, we choose $M = \sigma_x$ for convenience. As for the particle-hole symmetry, we adopt the principles in Ref. \cite{fang2015tridirac} that $P^2=1$ and $P$ commutes with spatial operations, \textit{i.e.}, $[P, M] = [P, e^{i\theta J_z}] = 0$, which fixes $P = \sigma_x\mcK$. We note that in the bosonic case the chiral symmetry becomes $S = TP = 1$, forcing the whole Hamiltonian to vanish. Hence, in the main text we only consider fermionic case, where the projected angular momentum quantum number $m$ is denoted as $n/2$.

\subsection{Details of surface symmetry analysis}

In Table \ref{tableS:surfacek.p_allowedterms}, we analyze the allowed terms in Eq. (\ref{eq:k.p_general}) by various symmetries involving $T$, $P$, $M$ and rotational symmetries from constraints Eqs. (\ref{eq:k.p_sym_general}), always assuming that the three-fold rotational symmetry $C_3$ is present (noting that $C_3 = \mathbb{I}$ up to a phase factor). We first focus on the discrete symmetries. To protect cubic dispersion without enforcing full isotropy, we need a combination of symmetries under which all allowed classes of terms have at least 3rd order in dispersion as indicated in the last row of Table \ref{tableS:surfacek.p_allowedterms}. These classes of terms include all terms with $n_+ - n_-$ being odd multiples of 3, as well as $ik_+^{n_+}k_-^{n_-}\sigma_{0,z} + H.c.$ with $n_+ - n_-$ being even multiples of 3, noting that in the latter type of terms the lowest order is 6 because all lower-order terms have $n_+ = n_-$ and thus are self-canceled by Hermiticity enforcement (``$+H.c.$''). An $SC_2$ symmetry (combined with $C_3$) is sufficient to satisfy the condition; alternatively, the combination $\{T, P\}$, $\{T, PM\}$, $\{P, M\}$ or $\{PM, C_2\}$ (together with $C_3$) also satisfy the condition to protect cubic band dispersion.

For the condition of cubic touching with quadratic bending allowed, the searching logic is similar -- the goal is to keep the lowest touching order of allowed terms above or at 3 as indicated in the second-to-last row of Table \ref{tableS:surfacek.p_allowedterms}. Compared to the cubic dispersion condition, one more class of terms is allowed, which is $k_+^{n_+}k_-^{n_-}\sigma_0 + H.c.$ with even $n_+ - n_-$. Two extra symmetry combinations protect this kind of cubic touching: $\{T\}$ and $\{M, C_2\}$ (together with $C_3$). Furthermore, the cubic touching can be broken down to three Dirac cones by also allowing $(i)k_+^{n_+}k_-^{n_-}\sigma_+ + H.c.$ with even $n_+ - n_-$ but disallowing all $\sigma_z$ components. Either $S$ or $TC_2$ (together with $C_3$) protects such structure, with one model example shown in a later section. All these have been summarized in Table \ref{table:sym_summary} in the main text.

\begin{table}[b]
\caption{Allowed terms of surface Hamiltonian under each individual symmetry, and the lowest allowed orders in band touching $E_\kk^+ - E_\kk^-$ and dispersions $E_\kk^\pm - E_\0^\pm$ (near $\Gamma$) in the presence of the individual classes of terms under the $C_3$ symmetry constraint that $n_+ \equiv n_-$ (mod 3). ``$-$'' means that the term does not open a gap.}
\begingroup
\newcommand{\ct}[1]{{\centering #1\par}}
\newcommand{\lc}[1]{
	\begingroup
	\setbox0=\hbox{#1}
	{\centering \raisebox{-\dimexpr\ht0-\dp0}{\large#1} \par}
	\endgroup
}
\begin{tabular}{|p{0.1\textwidth}|p{0.065\textwidth}|p{0.065\textwidth}|p{0.065\textwidth}|p{0.065\textwidth}|p{0.065\textwidth}|p{0.065\textwidth}|p{0.065\textwidth}|p{0.065\textwidth}|p{0.065\textwidth}|p{0.065\textwidth}|p{0.065\textwidth}|p{0.065\textwidth}|}
	\hline
	\ct{term}
	& \multicolumn{2}{c|}{$k_+^{n_+} k_-^{n_-}\sigma_0 \!+\! H.c.$}
	& \multicolumn{2}{c|}{$ik_+^{n_+} k_-^{n_-}\sigma_0 \!+\! H.c.$}
	& \multicolumn{2}{c|}{$k_+^{n_+} k_-^{n_-}\sigma_z \!+\! H.c.$}
	& \multicolumn{2}{c|}{$ik_+^{n_+} k_-^{n_-}\sigma_z \!+\! H.c.$}
	& \multicolumn{2}{c|}{$k_+^{n_+} k_-^{n_-}\sigma_+ \!+\! H.c.$}
	& \multicolumn{2}{c|}{$ik_+^{n_+} k_-^{n_-}\sigma_+ \!+\! H.c.$}
	\\\hline
	\ct{$n_+ - n_-$} & \ct{even} & \ct{odd} & \ct{even} & \ct{odd} & \ct{even} & \ct{odd} & \ct{even} & \ct{odd} & \ct{even} & \ct{odd} & \ct{even} & \ct{odd}  \\\hline
	\ct{$T = i\sigma_y\mcK$} & \ct{\Checkmark} & \ct{\XSolid} & \ct{\Checkmark} & \ct{\XSolid} & \ct{\XSolid} & \ct{\Checkmark} & \ct{\XSolid} & \ct{\Checkmark} & \ct{\XSolid} & \ct{\Checkmark} & \ct{\XSolid} & \ct{\Checkmark}  \\\hline
    \ct{$P = \sigma_x\mcK$} & \ct{\XSolid} & \ct{\Checkmark} & \ct{\XSolid} & \ct{\Checkmark} & \ct{\Checkmark} & \ct{\XSolid} & \ct{\Checkmark} & \ct{\XSolid} & \ct{\XSolid} & \ct{\Checkmark} & \ct{\XSolid} & \ct{\Checkmark}  \\\hline
	\ct{$S = TP$} & \ct{\XSolid} & \ct{\XSolid} & \ct{\XSolid} & \ct{\XSolid} & \ct{\XSolid} & \ct{\XSolid} & \ct{\XSolid} & \ct{\XSolid} & \ct{\Checkmark} & \ct{\Checkmark} & \ct{\Checkmark} & \ct{\Checkmark}  \\\hline
    \ct{$M = \sigma_x$} & \ct{\Checkmark} & \ct{\Checkmark} & \ct{\XSolid} & \ct{\XSolid} & \ct{\XSolid} & \ct{\XSolid} & \ct{\Checkmark} & \ct{\Checkmark} & \ct{\Checkmark} & \ct{\Checkmark} & \ct{\XSolid} & \ct{\XSolid}  \\\hline
    \ct{$TM$} & \ct{\Checkmark} & \ct{\XSolid} & \ct{\XSolid} & \ct{\Checkmark} & \ct{\Checkmark} & \ct{\XSolid} & \ct{\XSolid} & \ct{\Checkmark} & \ct{\XSolid} & \ct{\Checkmark} & \ct{\Checkmark} & \ct{\XSolid}  \\\hline
    \ct{$PM$} & \ct{\XSolid} & \ct{\Checkmark} & \ct{\Checkmark} & \ct{\XSolid} & \ct{\XSolid} & \ct{\Checkmark} & \ct{\Checkmark} & \ct{\XSolid} & \ct{\XSolid} & \ct{\Checkmark} & \ct{\Checkmark} & \ct{\XSolid}  \\\hline
    \ct{$SM$} & \ct{\XSolid} & \ct{\XSolid} & \ct{\Checkmark} & \ct{\Checkmark} & \ct{\Checkmark} & \ct{\Checkmark} & \ct{\XSolid} & \ct{\XSolid} & \ct{\Checkmark} & \ct{\Checkmark} & \ct{\XSolid} & \ct{\XSolid}  \\\hline
	\ct{$C_2 = \sigma_z$} & \ct{\Checkmark} & \ct{\XSolid} & \ct{\Checkmark} & \ct{\XSolid} & \ct{\Checkmark} & \ct{\XSolid} & \ct{\Checkmark} & \ct{\XSolid} & \ct{\XSolid} & \ct{\Checkmark} & \ct{\XSolid} & \ct{\Checkmark}  \\\hline
	\ct{$TC_2$} & \ct{\Checkmark} & \ct{\Checkmark} & \ct{\Checkmark} & \ct{\Checkmark} & \ct{\XSolid} & \ct{\XSolid} & \ct{\XSolid} & \ct{\XSolid} & \ct{\Checkmark} & \ct{\Checkmark} & \ct{\Checkmark} & \ct{\Checkmark}  \\\hline
    \ct{$PC_2$} & \ct{\XSolid} & \ct{\XSolid} & \ct{\XSolid} & \ct{\XSolid} & \ct{\Checkmark} & \ct{\Checkmark} & \ct{\Checkmark} & \ct{\Checkmark} & \ct{\Checkmark} & \ct{\Checkmark} & \ct{\Checkmark} & \ct{\Checkmark}  \\\hline
	\ct{$SC_2$} & \ct{\XSolid} & \ct{\Checkmark} & \ct{\XSolid} & \ct{\Checkmark} & \ct{\XSolid} & \ct{\Checkmark} & \ct{\XSolid} & \ct{\Checkmark} & \ct{\XSolid} & \ct{\Checkmark} & \ct{\XSolid} & \ct{\Checkmark}  \\\hline
    \ct{$J_z = n\sigma_z/2$} & \ct{\Checkmark} & \ct{\XSolid} & \ct{\XSolid} & \ct{\XSolid} & \ct{\Checkmark} & \ct{\XSolid} & \ct{\XSolid} & \ct{\XSolid} & \ct{\XSolid} & \ct{\Checkmark} & \ct{\XSolid} & \ct{\Checkmark}  \\\hline
	lowest order in touching & \lc{$-$} & \lc{$-$} & \lc{$-$} & \lc{$-$} & \lc{0} & \lc{3} & \lc{6} & \lc{3} & \lc{0} & \lc{3} & \lc{0} & \lc{3}  \\\hline
	lowest order in dispersion & \lc{2} & \lc{3} & \lc{6} & \lc{3} & \lc{2} & \lc{3} & \lc{6} & \lc{3} & \lc{2} & \lc{3} & \lc{2} & \lc{3}  \\\hline
\end{tabular}
\endgroup
\label{tableS:surfacek.p_allowedterms}
\end{table}

Now we consider the presence of the continuous rotational symmetry generated by $J_z$, which ensures isotropy and suppresses hexagonal warping. Because $J_z$ forces $n_+ = n_-$ in all $\sigma_0$ and $\sigma_z$ terms, it rules out all of $ik_+^{n_+}k_-^{n_-}\sigma_{0,z} + H.c.$. To protect isotropic cubic dispersion, we only need to further suppress $k_+^{n_+}k_-^{n_-}\sigma_{0,z}$ with even $n_+-n_-$, which can be achieved by $S$ or $PM$. For isotropic cubic touching with quadratic bending, we can allow $k_+^{n_+}k_-^{n_-}\sigma_0 + H.c.$ with even $n_+-n_-$ while still disallowing $k_+^{n_+}k_-^{n_-}\sigma_z + H.c.$ with even $n_+-n_-$. This can be achieved by $T$ or $M$. These are also summarized in Table \ref{table:sym_summary} in the main text.

\subsection{Symmetry fixing of bulk Hamiltonian}

We search for bulk Hamiltonian terms in our prism lattice model allowed by symmetries (including the particle-hole symmetry $P$)
\begin{equation}
\begin{array}{c}
    TH(\kk)T^{-1} = H(-\kk), \quad
    C_3H(\kk)C_3^{-1} = H(\mcR_{120^\circ}^z\kk), \quad
    C_2H(k_x, k_y, k_z)C_2^{-1} = H(-k_x, -k_y, k_z),  \\\addlinespace
    MH(k_x, k_y, k_z)M^{-1} = H(k_x, -k_y, k_z), \quad
    IH(\kk)I^{-1} = H(-\kk), \quad
    PH(\kk)P^{-1} = -H(-\kk),
\end{array}
\end{equation}
where $T$, $C_3$, $C_2$, $M$ and $I$ are given by Eq. (\ref{eq:bulk_syms}) in the main text and the form of $P$ is to be determined. We first note that $K = IT = i\tau^x\sigma^y\mcK$ preserves the crystal momentum, thus restricts the pseudospin sector of all symmetry-allowed Hamiltonian terms within those matrices commuting with $i\tau^x\sigma^y\mcK$. Besides the identity matrix, there are only 5 linearly independent matrices that satisfy this condition, which are the 4 Dirac matrices defined in Eq. (\ref{eq:Gamma_defs}), and their product $\Gamma^5 = \Gamma^1\Gamma^2\Gamma^3\Gamma^0 = \tau^z\sigma^z$. Next, we determine the bulk form of $P$ from the commutation algebra identified from the general surface band symmetry analysis (see Eq. (\ref{eq:surface_syms_general}) in the main text, and note that in surface representation $C_2 = e^{-i\pi J_z}$, which is the same as $\sigma_z$ up to a phase factor): $\{TP, T\} = \{TP, M\} = [TP, C_2] = 0$. ($TP$ is known as the chiral symmetry, which we sometimes denote as $S$.) The only candidates of $TP$ satisfying this algebra are $\sigma^z$, $\tau^x\sigma^z$ and $\tau^z\sigma^z$. Since $TP$ also preserves the momentum, it restricts the pseudospin part of Hamiltonian terms to those anticommuting with it. If $TP$ were set to $\sigma^z$ or $\tau^x\sigma^z$, then 3 out of the 5 Dirac matrices ($\Gamma^i$ for $i = 0, 1, 2, 3, 5$) would be banned from the Hamiltonian by the anticommutation condition, which would overconstrain our system. Hence, we finally set $TP = \tau^z\sigma^z$, \textit{i.e.} $P = \tau^z\sigma^x\mcK$, which only bans $\Gamma^5$ (and the identity matrix) from the bulk Hamiltonian. As for the momentum-space part, we consider up to second neighbor hoppings in both $z$ and in-plane directions. Note that $C_3$ symmetry does not act on pseudospins, thus the $C_3$-respecting candidates include the functions $\cos k_z$, $\sin k_z$, $c_{1,2}(k_x, k_y)$, $s_{1,2}(k_x, k_y)$ and their products. ($c_{1,2}$ and $s_{1,2}$ are defined in Eq. (\ref{eq:c_s}) in the main text.)

\begin{table}
\caption{Sign picked up by each candidate factor of bulk Hamiltonian under various symmetry transforms.}
\begingroup
\setlength{\tabcolsep}{6pt}
\centering
\begin{tabular}{|c|c|ccccc|cccccc|}
	\hline
    symmetry\textbackslash term & $H_0(\kk)$ & $\Gamma^0$ & $\Gamma^1$ & $\Gamma^2$ & $\Gamma^3$ & $\Gamma^5$ &
    $\cos k_z$ & $\sin k_z$ & $c_1(k_x, k_y)$ & $s_1(k_x, k_y)$ & $c_2(k_x, k_y)$ & $s_2(k_x, k_y)$ \\\hline
    $T$ & $+$ & $+$ & $-$ & $-$ & $-$ & $-$  &  $+$ & $-$ & $+$ & $-$ & $+$ & $-$ \\
    $P$ & $-$ & $-$ & $+$ & $+$ & $+$ & $-$  &  $+$ & $-$ & $+$ & $-$ & $+$ & $-$ \\
    $S = TP$ & $-$ & $-$ & $-$ & $-$ & $-$ & $+$  &  $+$ & $+$ & $+$ & $+$ & $+$ & $+$ \\
    $C_2$ & $+$ & $+$ & $-$ & $-$ & $+$ & $+$  &  $+$ & $+$ & $+$ & $-$ & $+$ & $-$ \\
    $C_3$ & $+$ & $+$ & $+$ & $+$ & $+$ & $+$  &  $+$ & $+$ & $+$ & $+$ & $+$ & $+$ \\
    $M$ & $+$ & $+$ & $+$ & $-$ & $+$ & $-$  &  $+$ & $+$ & $+$ & $+$ & $+$ & $-$ \\
    $I$ & $+$ & $+$ & $-$ & $-$ & $-$ & $-$  &  $+$ & $-$ & $+$ & $-$ & $+$ & $-$ \\
    $K = IT$ & $+$ & $+$ & $+$ & $+$ & $+$ & $+$  &  $+$ & $+$ & $+$ & $+$ & $+$ & $+$ \\\hline
\end{tabular}
\endgroup
\label{tableS:bulk_table1}
\end{table}

\begin{table}
\caption{Symmetry breaking effect of every candidate Hamiltonian term: for each combination of Dirac matrices and momentum-space functions, the corresponding entry in this table shows which one(s) of $T$, $P$, $C_2$, $M$, $K$ it breaks. Note that all these symmetries have eigenvalues $\pm 1$. Hence, an entry showing ``$A$, $B$, $C$'' means that the corresponding candidate Hamiltonian term breaks $A$, $B$, $C$ and $ABC$ but preserves $AB$, $AC$ and $BC$. Also note that all terms respect $C_3$ symmetry, and that multiplying any term with $\cos k_z$ and/or $c_{1,2}(k_x, k_y)$ does not change which symmetries the term breaks.}
\begingroup
\setlength{\tabcolsep}{6pt}
\centering
\begin{tabular}{|c|cccccc|}
    \hline
    & $1$ & $\sin k_z$ & $s_1(k_x,k_y)$ & $s_2(k_x,k_y)$ & $s_1(k_x,k_y)\sin k_z$ & $s_2(k_x,k_y)\sin k_z$  \\\hline
    $1$ & $P$ & $T$ & $T$, $C_2$ & $T$, $C_2$, $M$ & $P$, $C_2$ & $P$, $C_2$, $M$  \\
    $\Gamma^0$ & None & $T$, $P$ & $T$, $P$, $C_2$ & $T$, $P$, $C_2$, $M$ & $C_2$ & $C_2$, $M$  \\
    $\Gamma^1$ & $T$, $P$, $C_2$ & $C_2$ & None & $M$ & $T$, $P$ & $T$, $P$, $M$  \\
    $\Gamma^2$ & $T$, $P$, $C_2$, $M$ & $C_2$, $M$ & $M$ & None & $T$, $P$, $M$ & $T$, $P$  \\
    $\Gamma^3$ & $T$, $P$ & None & $C_2$ & $C_2$, $M$ & $T$, $P$, $C_2$ & $T$, $P$, $C_2$, $M$  \\
    $\Gamma^5$ & $T$, $M$ & $P$, $M$ & $P$, $C_2$, $M$ & $P$, $C_2$ & $T$, $C_2$, $M$ & $T$, $C_2$  \\\hline
    $i\Gamma^0\Gamma^1$ & $P$, $C_2$, $K$ & $T$, $C_2$, $K$ & $T$, $K$ & $T$, $M$, $K$ & $P$, $K$ & $P$, $M$, $K$  \\
    $i\Gamma^0\Gamma^2$ & $P$, $C_2$, $M$, $K$ & $T$, $C_2$, $M$, $K$ & $T$, $M$, $K$ & $T$, $K$ & $P$, $M$, $K$ & $P$, $K$  \\
    $i\Gamma^0\Gamma^3$ & $P$, $K$ & $T$, $K$ & $T$, $C_2$, $K$ & $T$, $C_2$, $M$, $K$ & $P$, $C_2$, $K$ & $P$, $C_2$, $M$, $K$  \\
    $i\Gamma^0\Gamma^5$ & $M$, $K$ & $T$, $P$, $M$, $K$ & all & $T$, $P$, $C_2$, $K$ & $C_2$, $M$, $K$ & $C_2$, $K$  \\
    $i\Gamma^1\Gamma^5$ & all & $C_2$, $M$, $K$ & $M$, $K$ & $K$ & $T$, $P$, $M$, $K$ & $T$, $P$, $K$  \\
    $i\Gamma^2\Gamma^5$ & $T$, $P$, $C_2$, $K$ & $C_2$, $K$ & $K$ & $M$, $K$ & $T$, $P$, $K$ & $T$, $P$, $M$, $K$  \\
    $i\Gamma^3\Gamma^5$ & $T$, $P$, $M$, $K$ & $M$, $K$ & $C_2$, $M$, $K$ & $C_2$, $K$ & all & $T$, $P$, $C_2$, $K$  \\
    $i\Gamma^1\Gamma^2$ & $T$, $M$, $K$ & $P$, $M$, $K$ & $P$, $C_2$, $M$, $K$ & $P$, $C_2$, $K$ & $T$, $C_2$, $M$, $K$ & $T$, $C_2$, $K$  \\
    $i\Gamma^1\Gamma^3$ & $T$, $C_2$, $K$ & $P$, $C_2$, $K$ & $P$, $K$ & $P$, $M$, $K$ & $T$, $K$ & $T$, $M$, $K$  \\
    $i\Gamma^2\Gamma^3$ & $T$, $C_2$, $M$, $K$ & $P$, $C_2$, $M$, $K$ & $P$, $M$, $K$ & $P$, $K$ & $T$, $M$, $K$ & $T$, $K$  \\\hline
\end{tabular}
\endgroup
\label{tableS:bulk_table2}
\end{table}

We analyze the symmetry transform of each pseudospin and momentum candidate in Table \ref{tableS:bulk_table1}. Based on it, the symmetries broken by each combination of them are analyzed and presented in Table \ref{tableS:bulk_table2}, 
in terms of which out of the generators $T$, $P$, $C_2$, $M$ and $K$ the candidate violates (note that all candidates respect $C_3$). We see that the terms that respect all symmetries include $\Gamma^0$, $\Gamma^1s_1(k_x, k_y)$, $\Gamma^2s_2(k_x, k_y)$, $\Gamma^3\sin k_z$ and the product of each individual of them with $\cos k_z$, $c_1(k_x, k_y)$ or $c_2(k_x, k_y)$. To make our model simple, we only include the terms $\Gamma^0$, $\Gamma^0\cos k_z$, $\Gamma^0 c_1(k_x, k_y)$, $\Gamma^1s_1(k_x, k_y)$, $\Gamma^2s_2(k_x, k_y)$ and $\Gamma^3\sin k_z$ in the bulk Hamiltonian $H_0(\kk)$, as presented in Eqs. (\ref{eq:bulkH})-(\ref{eq:h}) in the main text, where we make the coefficients of $\Gamma^0\cos k_z$ and $\Gamma^3\sin k_z$ equal in order to cancel distant vertical hoppings (\textit{i.e.} so that they are both given by the single spin-independent vertical interlayer coupling $t''$ illustrated in Fig. \ref{fig:model}(a) in the main text). The skew hopping term $\Gamma^0 c_1(k_x, k_y)$ is necessary to realize the nontrivial topology, which will be shown in the next section. In addition, we see from Table \ref{tableS:bulk_table2} that the only nontrivial terms that violate $P$ and respect all other symmetries are $\cos k_z$, $c_{1,2}(k_x, k_y)$ and their products. $\cos k_z$ involves interlayer hopping between the same sublattice, which we ignore in our model. Hence, we fix our form of $H_P(\kk)$ to Eq. (\ref{eq:HP}) in the main text, which captures the physically inevitable in-plane hoppings.

\section{Additional prism-model results and topological phases}

\begin{figure}
	\centering
	\includegraphics[width=\textwidth]{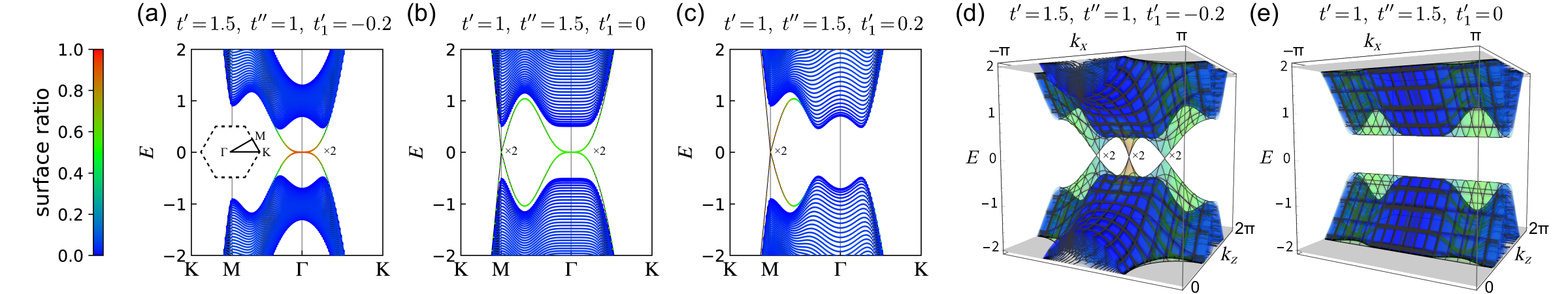}
	\caption{(a)-(c) Band structure of a 50-bilayer $xy$-slab with the same geometry as studied in Fig. \ref{fig:surfacestates3Dplot} of the main text, with $t_1^R = 1$, $t_2^R = 0.2$, $t_1 = t_2 = 0$ and other parameters shown on the top of each panel. The bands are plotted along the solid high-symmetry lines shown in the inset of (a). Note that (a) has the same parameters as for the 3D plot shown in Fig. \ref{fig:surfacestates3Dplot}(a). (d)-(e) 3D plots of the band structure of a $xz$-slab with 48 unit cells along the $y$ direction, with $t_1^R = 1$, $t_2^R = 0.2$, $t_1 = t_2 = 0$ and other parameters shown on the top of each panel. All panels use the color scale shown on the left to illustrate the bulk/surface nature of bands, same as in Fig. \ref{fig:surfacestates3Dplot} of the main text.}
	\label{figS:surfacestates}
\end{figure}

In Figs. \ref{figS:surfacestates} (a)-(c), we present the $xy$-slab band structure under three distinct sets of model parameters in $H_0(\kk)$. We see that surface band touching can occur at $\Gamma$, $\rm M$ or both points, and that the band touching at $\rm M$ is linear rather than cubic, due to the lack of $C_3$ symmetry at a single $\rm M$ point. The presence and absence of surface band touching can be understood by examining the effective 1D models at $\Gamma$ and $\rm M$:
\begin{subequations}
\begin{gather}
    \label{eq:eff_SSH_Gamma}
	H_0(\Gamma, k_z) = (t' + 6t_1' + t''\cos k_z) \, \tau^x + t''\sin k_z \, \tau^y, \\\addlinespace
	H_0({\rm M}, k_z) = (t' - 2t_1' + t''\cos k_z) \, \tau^x + t''\sin k_z \, \tau^y,
\end{gather}
\end{subequations}
both of which are two spin-degenerate copies of SSH models \cite{su1979solitons} with presence and absence of zero edge modes controlled by the 1D winding number: surface band touching exists at $\Gamma$ ($\rm M$) if and only if $|t''| > |t'+6t_1'|$ ($|t''| > |t'-2t_1'|)$. We see that a nonzero skew hopping $t_1'$ is crucial to make the 1D winding numbers at $\Gamma$ and $\rm M$ different, and thus to allow the surface band touching point to only occur at $\Gamma$ (or only occur at the $\rm M$ points).

We identify the topological phases these three results belong to. If we only consider the local symmetries $T$ and $P$ of $H_0(\kk)$, then the algebra $T^2 = -1$ and $P^2 = 1$ places the system in class DIII according to the Altland-Zirnbauer classification scheme \cite{zirnbauer1996riemannian, altland1997nonstandard, schnyder2008classification, kitaev2009periodic}, giving a protected $\bbZ$ topological invariant corresponding to the 3D winding number \cite{schnyder2008classification}. The cubic touching of surface bands in Fig. \ref{figS:surfacestates}(a) indicates winding number 3, which is consistent with the presence of 3 Dirac cones in $xz$ surface bands illustrated in Fig. \ref{figS:surfacestates} (d). For the model parameters shown in Figs. \ref{figS:surfacestates} (b), (c), because there are 3 inequivalent $\rm M$ points in the BZ related by $C_3$ rotation, there are 3 Dirac cones in the surface bands each at one $\rm M$ point, totally contributing a topological charge with absolute value $3$. We argue that when band touching exists at both $\Gamma$ and $\rm M$, their topological charges cancel, leading to winding number 0 -- because Fig. \ref{figS:surfacestates}(e) shows that the surface bands of an $xz$-slab do not touch under the same parameters as in (b). Therefore, the system in (b) is ``trivial'' with respect to DIII topology -- breaking of translational symmetry that folds $\rm M$ points to $\Gamma$ can open gaps in the band touching. We also comment that when the particle-hole symmetry is broken, the Altland-Zirnbauer class becomes AII, with the topological invariant reduced to $\bbZ_2$. In this sense, both (a) and (c) have odd $\bbZ_2$ so they are ``topologically equivalent'' (after breaking $P$). 
Hence, the non-triviality of Fig. \ref{figS:surfacestates} (b) and topological distinction between (a) and (c) with broken $P$ have to be understood under classification regimes beyond Altland-Zirnbauer.

In the presence of $P$, we argue that the presence and absence of band touching at individual $\rm M$ and $\Gamma$ points are controlled by 1D weak $\bbZ_2$ topological indices -- the topological protection requires translational symmetry not to be broken along either of the two in-plane dimensions. The bulk system has time-reversal symmetry $T = i\sigma^y\mcK$ and spatial inversion $I = \tau^x$, thus the 1D weak $\bbZ_2$ invariant is determined by the product of inversion eigenvalues at two inversion-invariant momenta below the bulk gap \cite{fu2007topological, teo2008surface, hughes2011inversionsymmetric}:
\begin{equation}
    (-1)^{\bbZ_2(\overline\Gamma)} = \xi(\Gamma) \xi({\rm Z}), \quad (-1)^{\bbZ_2(\overline{\rm M})} = \xi({\rm M}) \xi({\rm M_Z}),
\end{equation}
where we have re-denoted the 2D high-symmetry points by adding an overline to distinguish from 3D high-symmetry points (see Fig. \ref{fig:model}(c) for positions of $\Gamma$, $\rm Z$, $\rm M$ and $\rm M_Z$ in the 3D BZ). These indices are exactly the parities of the 1D winding numbers of effective SSH models identified earlier in this section.

While allowing for an extra $P$-breaking term, the 1D weak topology is no longer protected \cite{altland1997nonstandard}, thus the state in Fig. \ref{figS:surfacestates}(b) becomes topologically trivial again, and the states in Figs. \ref{figS:surfacestates} (a) and (c) become topologically equivalent with respect to both strong and weak topologies.
Here we argue that (a) and (c) still belong to distinct topological crystalline phases protected by the mirror symmetries \cite{hsieh2012topological}. The mirror symmetry $MH(k_x, k_y, k_z)M^{-1} = H(k_x, -k_y, k_z)$ ($M = \sigma^x$) defines a mirror Chern number, which is the Chern number of the effective 2D system in the mirror-invariant plane $k_y = 0$ in the $\sigma^x = 1$ sector,
\begin{equation}
\begin{aligned}
    H_M(k_x, k_z) &= \!\lp. H(k_x, 0, k_z) \rp|_{\sigma^x = 1} \\\addlinespace
    &= t_1c_1(k_x, 0) + t_2c_2(k_x, 0) + \bigl( t' + t_1'c_1(k_x, 0) + t''\cos k_z \bigr)\,\tau^x + t''\sin k_z\,\tau^y + t_1^Rs_1(k_x, 0)\,\tau^z
\end{aligned}
\end{equation}
(note that $s_2(k_x, 0) = 0$). This mirror Chern number is 1, 0 and $-1$ in the systems shown in Figs. \ref{figS:surfacestates} (a), (b) and (c), respectively. Adding nonzero $t_1$ and $t_2$ does not change the mirror Chern number.
In addition, $C_2$ symmetry in our system gives rise to another mirror symmetry $M'H(k_x, k_y, k_z)M'^{-1} = H(-k_x, k_y, k_z)$, where $M' = C_2M = \sigma^y$ (up to a phase factor). The mirror Chern number associated with $M'$ is the Chern number of the effective 2D system in the $M'$-invariant plane $k_x = 0$ in the $\sigma^y = 1$ sector,
\begin{equation}
\begin{aligned}
    H_{M'}(k_y, k_z) &= \!\lp. H(0, k_y, k_z) \rp|_{\sigma^y = 1} \\\addlinespace
    &= t_1c_1(0, k_y) + t_2c_2(0, k_y) + \bigl( t' + t_1'c_1(0, k_y) + t''\cos k_z \bigr)\,\tau^x + t''\sin k_z\,\tau^y + t_2^Rs_2(0, k_y)\,\tau^z
\end{aligned}
\end{equation}
(note that $s_2(0, k_y) = 0$). This mirror Chern number is $-1$, 0 and 1 in the systems shown in Figs. \ref{figS:surfacestates} (a), (b) and (c), respectively. Other mirrors are related to $M$ or $M'$ by $C_3$ symmetry. We point out that the mirror Chern numbers themselves still do not fix whether the surface band touching is at $\Gamma$ or $M$ points.

\subsection{Surface density of states}

\begin{figure}
	\centering
	\includegraphics[width=\textwidth]{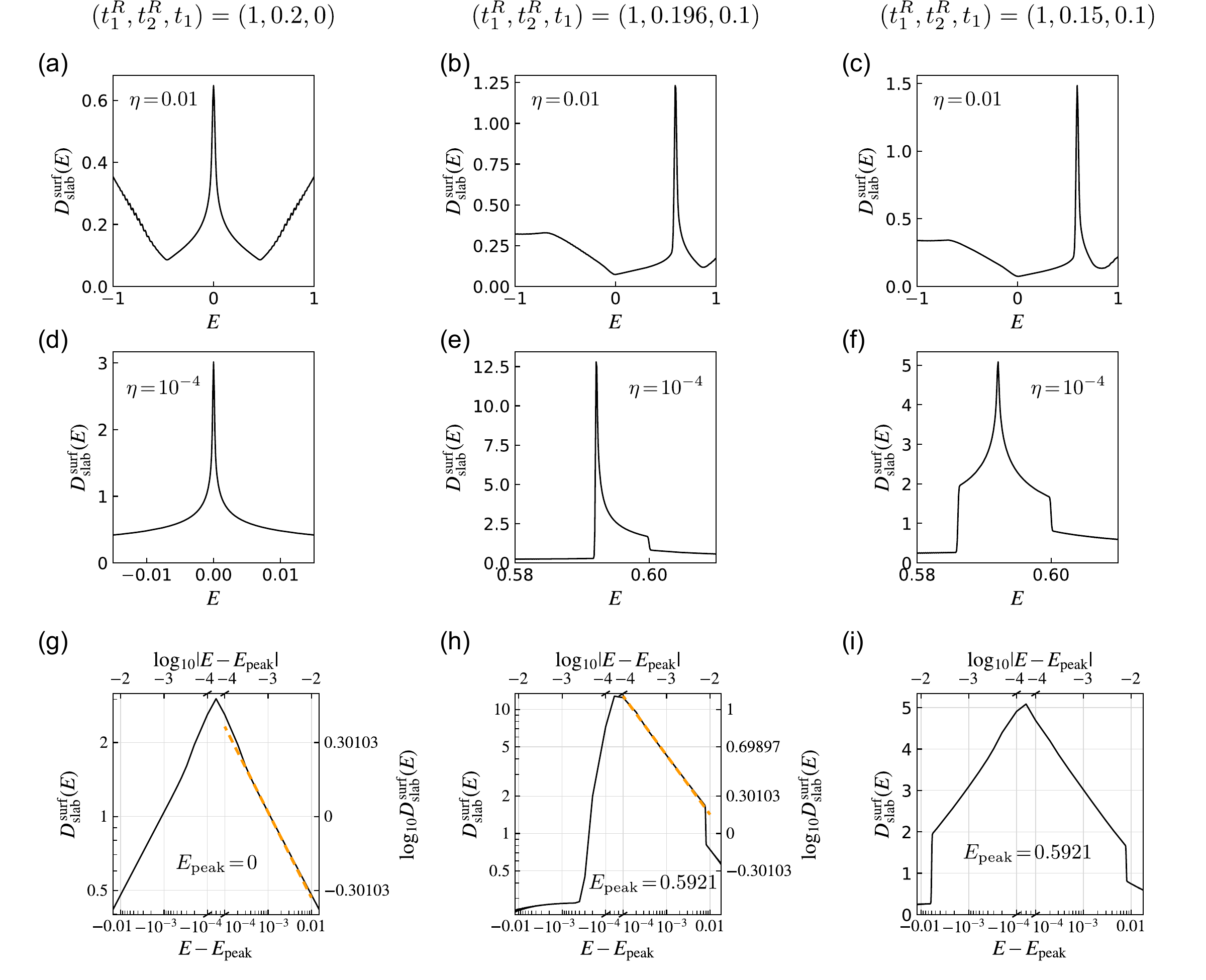}
	\caption{(a)-(c) The bottom surface DOS of the 50-cell slab studied in the main text, with Gaussian broadening $\eta = 0.01$, $t_1^R$, $t_2^R$ and $t_1$ marked on the top and other model parameters same as the system shown in Fig. \ref{fig:surfacestates3Dplot} of the main text. A $384 \times 384$ sample is taken in the 2D Brillouin zone. (d)-(f) The same as (a)-(c) except that here the Gaussian broadening $\eta = 10^{-4}$ and the Brillouin zone discretization is $3072 \times 3072$. (g)-(i) The same plots as (d)-(f) with logarithm curving in the energy relative to the marked peak energy $E_{\rm peak}$; in (g), (h), the DOS axis is also logarithm-curved. The orange dashed line fits the linear region in the log-log curve, with slope approximately $-1/3$ in (g) and $-1/2$ in (h), indicating DOS divergence $\sim E^{-1/3}$ ($\sim E^{-1/2}$). In (i), the linear regions in the graph of DOS vs. $\log_{10}|E-E_{\rm peak}|$ indicates logarithm divergence. The curvature near the peak is due to Gaussian broadening.}
	\label{figS:DOS}
\end{figure}

We compute the bottom surface DOS of the slab that contains $N_c = 50$ cells in the $z$-direction as we study in the main text, with a Gaussian broadening factor $\eta$.
The momentum-resolved bottom surface DOS includes contribution from both sublattices on the bottom unit cell layer, and is defined as
\begin{equation}
	D_{\rm slab}^{\rm surf}(\kk, E) = \sum_{n=1}^{4N_c} \bra{u_{\rm slab}^{n\kk}}\mathcal{P}_b\ket{u_{\rm slab}^{n\kk}} \Delta_\eta\bigl( E - E_{\rm slab}^n(\kk) \bigr),
\end{equation}
where $\ket{u_{\rm slab}^{n\kk}}$ is the $n$th normalized eigenstate of the slab Hamiltonian, $E_{\rm slab}^n(\kk)$ is the corresponding eigenvalue, $\mathcal{P}_b$ is the projection operator to the bottom unit cell, and $\Delta_\eta(\epsilon) = \frac{e^{-\epsilon^2/2\eta^2}}{\sqrt{2\pi}\eta}$ is the Gaussian distribution function. The total bottom surface DOS is $D_{\rm slab}^{\rm surf}(E) = \int\frac{d^2\kk}{(2\pi)^2} D_{\rm slab}^{\rm surf}(\kk, E)$.
Figs. \ref{figS:DOS} (a)-(c) show peaks in the broadened surface DOS, which respectively correspond to the cubic touching point, moat minimum and saddle points discussed in the main text. As shown in Figs. \ref{figS:DOS} (d)-(f), all these peaks increase as the broadening factor $\eta$ decreases, indicating singularity in the DOS.
The logarithm-curved plots in Figs. \ref{figS:DOS} (g)-(i) confirm that the singularities in (d), (e) and (f) are respectively $E^{-1/3}$, $E^{-1/2}$ and $\log E$, consistent with our main argument on the allowed types of singularities.

We note, however, that the $E^{-1/2}$ singularity is not obtained exactly at $t_2^R/t_1^R = 0.2$ as shown in Fig. \ref{fig:surfacestates3Dplot} (b), (c) in the main text, nor at the exact value $t_2^R/t_1^R = 1/3\sqrt{3} \approx 0.19245$ we identify for isotropic cubic touching at $\Gamma$ -- at the finite moat-minimum radius, higher-order components in the dispersion can also contribute to hexagonal warping. In particular, when $t_2 = 0$, the upper surface band dispersion is
\begin{equation}
\begin{aligned}
    E_+(k_x, k_y) &= t_1c_1(k_x, k_y) + \Abs{t_1^Rs_1(k_x, k_y) + it_2^Rs_2(k_x, k_y)}  \\\addlinespace
    &= 6t_1\lp( 1-\frac{k^2}{4} + \frac{k^4}{16} \rp) + \Abs{ -\frac{1}{4} t_1^R \lp( k^3 - \frac{k^5}{16} \rp) \cos 3\theta_k - \frac{3\sqrt{3}}{4}it_2^R \lp( k^3 - \frac{3k^5}{16} \rp) \sin 3\theta_k } + O(k^6)
\end{aligned}
\end{equation}
(see Eq. (\ref{eq:c_s}) in the main text for definitions of $c_j$, $s_j$), where $k = \sqrt{k_x^2 + k_y^2}$ and $\theta_k = \arctan(k_y/k_x)$. Let $t_2^R = (1+\epsilon)t_1^R/3\sqrt{3}$ deviate from the isotropic-touching limit by factor $1+\epsilon$ (i.e., let $\gamma = 1+\epsilon$ in Eq. (\ref{eq:mcH0_prism})). At radius $k$, the warping amplitude is about
\begin{equation}
    E_+(0, k) - E_+(k, 0) \approx \frac{t_1^R}{4} \lp( \epsilon k^3 - \frac{2+3\epsilon}{16} k^5 \rp),
\end{equation}
vanishing at $\epsilon = 2k^2/(16-3k^2)$.
The moat radius $k_m = 4t_1/t_1^R$, which is approximated by minimizing $E_+(k_m, 0) = 6t_1 - 3t_1k_m^2/2 + t_1^Rk_m^3/4 + O(k_m^4)$ to the leading order, takes the value $k_m = 0.4$ under our parameter value choice $t_1 = 0.1$, $t_1^R = 1$. At this radius, if $\epsilon = 0$, the warping will be $t_1^Rk_m^5/32 = 3.2\times 10^{-4}$, which exceeds the broadening factor $\eta$ we take in Figs. \ref{figS:DOS} (d)-(f). To suppress the warping at this radius, we need $\epsilon = 2k_m^2/(16-3k_m^2) \approx 0.02$, corresponding to $1.02t_1^R/3\sqrt{3} \approx 0.196$ as chosen in Figs. \ref{figS:DOS} (b), (e) and (h).

\subsection{Additional symmetry-breaking analysis of the prism lattice model}

\begin{figure}
	\centering
	\includegraphics[width=\textwidth]{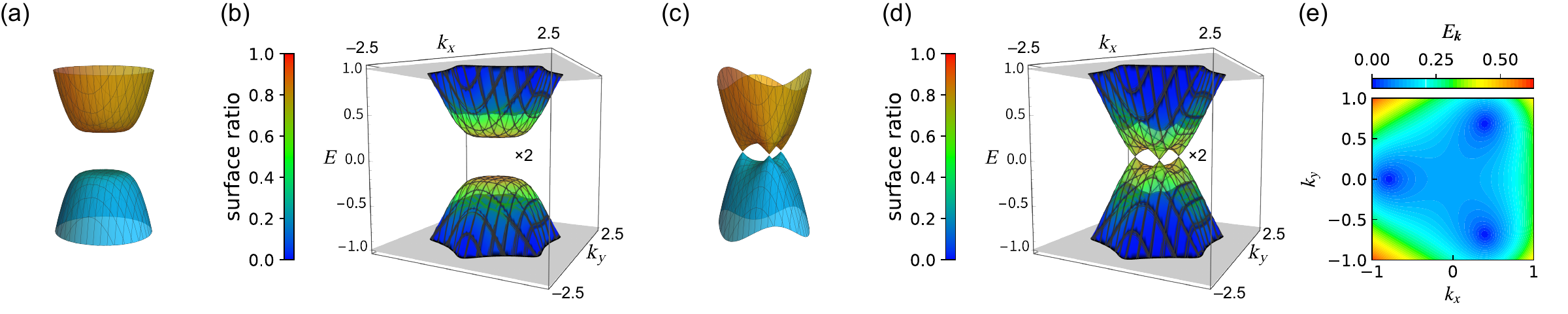}
	\caption{(a) Schematic of a gap opened from an isotropic cubic-touching structure. (b) The band structure of the $xy$-slab of our prism tight-binding model containing 50 unit cells along $z$ direction with the same parameters in $H_0(\kk)$ as studied in Fig. \ref{fig:surfacestates3Dplot}(a) in the main text, and one extra Hamiltonian term $H_{T,M} = \lambda_{T,M}\Gamma^5$ with parameter $\lambda_{T,M} = 0.25$. The color bar represents the bulk/surface nature of bands, as in Fig. \ref{fig:surfacestates3Dplot}. (c) Schematic of 3 Dirac cones split from a cubic touching point. (d) The slab band structure under the same $H_0$, but now instead of $H_{T,M}$, we include two extra Hamiltonian terms $H_{T,P}$ and $H_{C_2}$ with parameters $\lambda_{T,P} = \lambda_{C_2} = 0.25$. (e) The topography of the first surface band above the Dirac touching point in (d).}
	\label{figS:symbroken}
\end{figure}

We explore the symmetry protection of cubic band touching by selecting various symmetry breaking terms from Table \ref{tableS:bulk_table2} to add to $H_0(\kk)$ of our prism model. We first note that all terms containing factor $s_1(k_x, k_y)$ or $s_2(k_x, k_y)$ preserve the cubic touching even if they break all symmetries, because they are third order in $(k_x, k_y)$ at $\Gamma$ point thus can only have cubic correction to the surface $\kk\cdot\pp$ Hamiltonian $\mcH^{\rm surf}(\kk)$. Hence, in the rest of our analysis, we only consider terms that are independent of momentum or only contain factor $\sin k_z$ in momentum dependence. We first discuss the case where a single symmetry-breaking term violates every symmetry combination mentioned in Table \ref{table:sym_summary} in the main text, and allows for gap opening by introducing zeroth-order $\sigma_z$ component to the surface Hamiltonian (see Table \ref{tableS:surfacek.p_allowedterms}). A schematic band structure is shown in Fig. \ref{figS:symbroken}(a). We identify one such term, $H_{T,M} = \lambda_{T,M} \Gamma^5$, which breaks $T$ and $M$, and preserves $P$, $C_3$ and $C_2$ (thus it also preserves $TM$ but breaks $S = TP$). As Fig. \ref{figS:symbroken}(b) shows, this term opens a mass gap in the cubic-touching surface bands and flattens their dispersion to 6th order, which has DOS divergence $\sim E^{-2/3}$. However, the 6th-order dispersion is not symmetry-protected, but is rather a result of the specific $\kk$-independent form of $H_{T,M}$ we choose.

Next, we discuss cases where with the presence of some symmetry-breaking terms, the cubic dispersion and touching structure is still protected by remaining symmetries. To show that the cubic dispersion and touching can be protected by either $\{C_3, SC_2\}$ or $\{C_3, T, P\}$ (see Table \ref{table:sym_summary} in the main text), we introduce the terms $H_{T,P} = \lambda_{T,P}\Gamma^3$, $H_{C_2} = 2\lambda_{C_2}\Gamma^1\sin k_z$ and $H_{M,K} = i\lambda_{M,K}\Gamma^0\Gamma^5$, where the subscripts indicate which of the elementary symmetries the term breaks (see Table \ref{tableS:bulk_table2}). When only $H_{T,P}$ and $H_{M,K}$ are present, the cubic touching is protected by $\{C_3, SC_2\}$ (note that $S = TP$ so $H_{T,P}$ preserves $S$); when only $H_{C_2}$ and $H_{M,K}$ are present, the cubic touching is protected by $\{C_3, T, P\}$. However, if both $H_{T,P}$ and $H_{C_2}$ are present, or all the three terms are present, then the cubic touching is split into 3 Dirac cones (a schematic is shown in Fig. \ref{figS:symbroken}(c)), which are protected by $\{C_3, S\}$. Fig. \ref{figS:symbroken}(d) shows one specific case from our model results, where $\lambda_{T,P} = \lambda_{C_2} = 0.25$ and $\lambda_{M,K} = 0$. As illustrated in Fig. \ref{figS:symbroken}(e), the upper surface band has three minima, corresponding to the three Dirac cones, which interact to make a third-order VHS at $\kk = \0$. The third-order VHS has DOS divergence $\sim E^{-1/3}$, which again is not symmetry-protected -- perturbative quadratic terms that respect both $C_3$ and $S$ symmetries, \textit{e.g.} $\alpha(k_x^2+k_y^2)\sigma_x$, can split it into 3 normal saddle points, weakening the DOS divergence to $\sim -\ln E$.

Besides $\{C_3, SC_2\}$ and $\{C_3, T, P\}$, there are 3 other combinations that can protect the cubic dispersion and touching, as shown in Table \ref{table:sym_summary}. To address them, we consider two more symmetry-breaking terms: $H_{P,M} = 2\lambda_{P,M}\Gamma^5\sin k_z$ and $H_T = 2\lambda_T\sin k_z$. From symmetry argument, including any two of the three terms $H_{C_2}$, $H_{P,M}$ and $H_T$ preserves the cubic touching:
\begin{itemize}
    \item $H_{C_2} + H_{P,M}$ --- protected by $\{C_3, T, PM\}$;
    \item $H_{C_2} + H_T$ --- protected by $\{C_3, P, M\}$;
    \item $H_{P,M} + H_T$ --- protected by $\{C_6, PM\}$.
\end{itemize}
If all the three terms are present, then the system does not explicitly have any symmetry combination listed in Table \ref{table:sym_summary}. However, it turns out that the cubic band touching is \textit{still present} -- we note that $H_{C_2} = 2\tau^z\sigma^x\sin k_z$ and $H_{P,M} = 2\tau^z\sigma^z\sin k_z$, which means that regardless of their relative strength, $H_{C_2} + H_{P,M}$ can always be smoothly gauge-transformed to a single term $H_{C_2}$ (or $H_{P,M}$) via a spin gauge transformation, which does not change the effective 1D model at $\Gamma$ (note from Eq. (\ref{eq:eff_SSH_Gamma}) that the effective 1D model does not have spin splitting); and that the change in $H_0(\kk)$ caused by the gauge transformation is of cubic order in $(k_x, k_y)$, meaning that the cubic touching is preserved. This is one example where pure symmetry analysis is insufficient to predict band dispersion type because of gauge equivalence to systems with locally higher symmetry.

\subsection{An example of extended model}

\begin{figure}
    \centering
    \includegraphics[width = 0.75\textwidth]{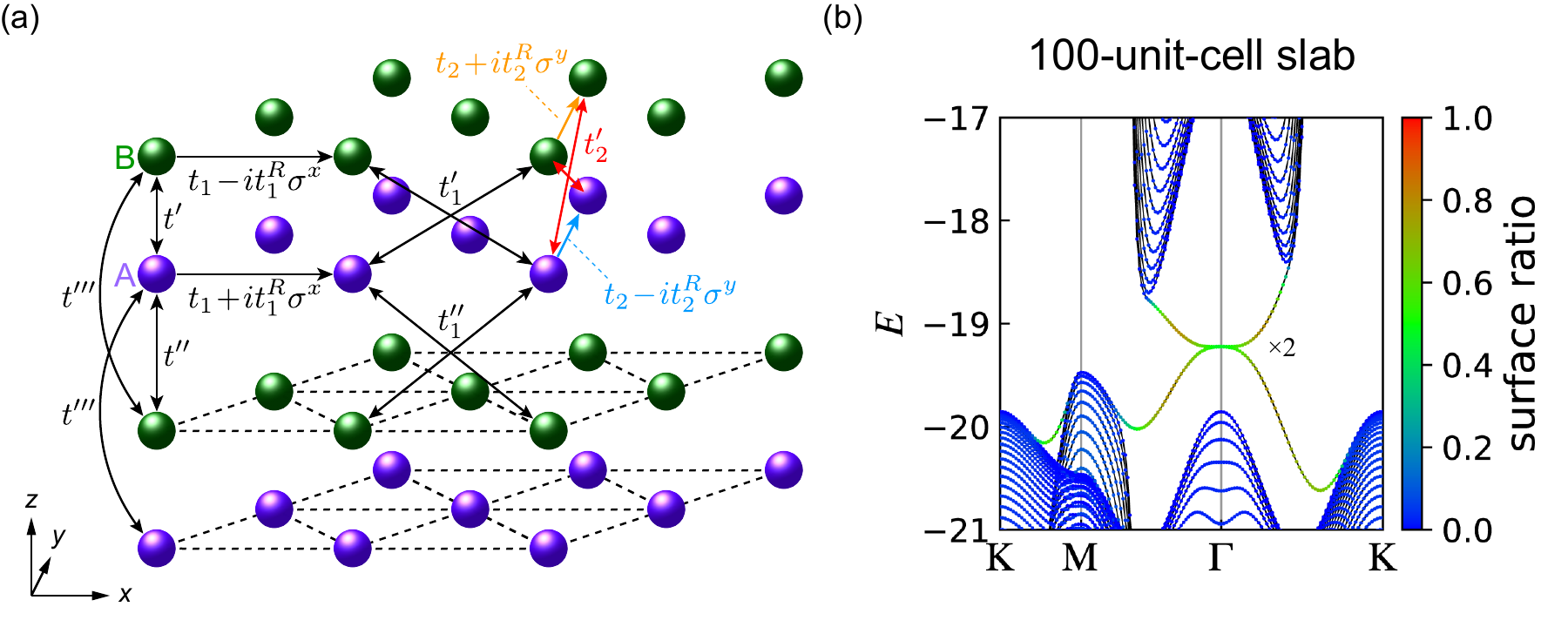}
    \caption{(a) A schematic of the extended prism lattice model with hoppings marked with arrows. (b) The partial band structure of a slab of the extended prism lattice model with 100 cells along $z$-axis. Here $t_1 = -4.78$, $t_2 = -0.45$, $t_1^R = 2.32$, $t_2^R = 0.44$, $t' = 19.5$, $t_1' = -4.77$, $t_2' = -0.69$, $t'' = -19.5$, $t_1'' = -0.15$ and $t''' = -9.35$. Like in the main text, the color scale reflects the bulk/surface nature of bands.}
    \label{figS:extended_model}
\end{figure}
As Fig. \ref{figS:extended_model}(a) shows, the extended tight-binding model on the prism lattice contains the nearest-layer skew hopping up to the second neighbor $t_2'$, the first skew hopping between the next-nearest neighboring layers $t_1''$ and the same-sublattice vertical hopping $t'''$, as well as other parameters already defined in the main text. All hoppings except $t_1^R$ and $t_2^R$ are spin-independent. All terms respect the symmetries $T$, $C_3$, $C_2$, $M$ and $I$, and only $t_1$, $t_2$ and $t'''$ violate the particle-hole symmetry $P$. Fig. \ref{figS:extended_model}(b) shows that by fine-tuning the parameters, the cubic-touching and singular points of the surface bands can be kept in the insulating gap even under $|t_1^R| < |t_1|$ and $|t_2^R| < |t_2|$.

\section{Routes to suppression of spin-independent hopping between $j_z = \pm 3/2$ states}

We use $p$ and $d$ atomic orbits to assemble the $j_z = \pm 3/2$ states. The $j_z = +3/2$ ($-3/2$) atomic state is a mixture of $\ket{p_{+1}\up}$, $\ket{d_{+1}\up}$ and $\ket{d_{+2}\down}$ ($\ket{p_{-1}\down}$, $\ket{d_{-1}\down}$ and $\ket{d_{-2}\up}$), where the subscript stands for the orbital magnetic quantum number and the arrow stands for the physical spin of the electron. Let $E_p$ and $E_d$ be the bare-atom non-relativistic energies of the $p$ and $d$ orbits, and include a crystal field effect which has an odd-in-$z$ component that couples $p$ and $d$ orbitals with opposite real amplitudes in the two sublattices ($\Delta_{pd}\tau^z$), and an even-in-$z$ component that splits the $d_{\pm1}$ and $d_{\pm2}$ energies with sublattice-independent amplitude $\Delta_d$.
Ref. \cite{shanavas2014theoretical} discusses the special case of $\Delta_d = 0$ in a different lattice model setting.
We also include an onsite spin-orbital coupling (SOC) term $\lambda\LL\cdot\SSS$. On both the basis $\{\ket{p_{+1}\up}, \ket{d_{+1}\up}, \ket{d_{+2}\down}\}$ and the basis $\{\ket{p_{-1}\down}, \ket{d_{-1}\down}, \ket{d_{-2}\up}\}$ on sublattice A, the total onsite Hamiltonian takes the same $3\times 3$ matrix form:
\begin{equation}
    H_{\rm onsite} = \begin{pmatrix}
        E_p + \frac{\lambda}{2} & \Delta_{pd} & 0  \\\addlinespace
        \Delta_{pd} & E_d + \frac{\lambda}{2} & \lambda  \\\addlinespace
        0 & \lambda & E_d-\lambda+\Delta_d
    \end{pmatrix}.
\end{equation}

Now we obtain the $j_z = \pm3/2$ orbitals by diagonalizing the onsite Hamiltonian under the assumption $|\Delta_{pd}|, |\Delta_d|, |\lambda| \ll E_d - E_p$. In the sectors $\ket{j_z = +3/2}$ and $\ket{j_z = -3/2}$, we get the approximate eigenvectors
\begin{subequations}
\begin{align}
    \Ket{\tilde p, +} \,&= \Ket{p_{+1}\up} - \eta\Ket{d_{+1}\up},  \\
    \Ket{\tilde d, +} \,&= \eta\cos\theta \Ket{p_{+1}\up} + \cos\theta \Ket{d_{+1}\up} + \sin\theta \Ket{d_{+2}\down},  \label{eqS:d,+}\\
    \Ket{\tilde d', +} \,&= -\eta\sin\theta \Ket{p_{+1}\up} - \sin\theta \Ket{d_{+1}\up} + \cos\theta \Ket{d_{+2}\down},
\end{align}
\end{subequations}
and
\begin{subequations}
\begin{align}
    \Ket{\tilde p, -} \,&= \Ket{p_{-1}\down} - \eta\Ket{d_{-1}\down},  \\
    \Ket{\tilde d, -} \,&= \eta\cos\theta \Ket{p_{-1}\down} + \cos\theta \Ket{d_{-1}\down} + \sin\theta \Ket{d_{-2}\up},  \label{eqS:d,-}\\
    \Ket{\tilde d', -} \,&= -\eta\sin\theta \Ket{p_{-1}\down} - \sin\theta \Ket{d_{-1}\down} + \cos\theta \Ket{d_{-2}\up},
\end{align}
\end{subequations}
respectively, where
\begin{equation}
    \eta = \frac{\Delta_{pd}}{E_d-E_p}, \quad
    \tan 2\theta = \frac{2\lambda}{\frac{3}{2}\lambda + \eta\Delta_{pd} - \Delta_d}.
    \label{eqS:eta}
\end{equation}
The corresponding eigenvalues in both sectors are
\begin{subequations}
\begin{align}
    E_{\tilde p} \;&= E_p + \frac{\lambda}{2} - \eta\Delta_{pd},  \\
    E_{\tilde d} \;&= E_d - \frac{\lambda}{4} + \frac{\eta\Delta_{pd} + \Delta_d}{2} + \sqrt{\lambda^2 + \lp( \frac{3}{4}\lambda + \frac{\eta\Delta_{pd} - \Delta_d}{2} \rp)^2},  \\
    E_{\tilde d'} \;&= E_d - \frac{\lambda}{4} + \frac{\eta\Delta_{pd} + \Delta_d}{2} - \sqrt{\lambda^2 + \lp( \frac{3}{4}\lambda + \frac{\eta\Delta_{pd} - \Delta_d}{2} \rp)^2}.
\end{align}
\end{subequations}

Next, we obtain the in-plane hopping amplitudes. We take hopping along a nearest-neighbor bond $\hat x = (1, 0, 0)$ as an example, only include spin-independent regular hoppings in the physical space and exclude the crystal-field- and SOC-induced hoppings.
We take hopping between $\tilde d$ states as an example, as the $\tilde d'$ states relates to $\tilde d$ by replacing $\theta$ with $\theta + \pi/2$, and spin-flip hopping between $\tilde p$ states vanish because $\tilde p$ states are spin-pure.
Within the same $j_z$ sector, $p_{\pm1}$ cannot hop to $d_{\pm1}$ states because of $z$-parity mismatch, and physical spin-flip hoppings are also absent, making the hopping matrix diagonal:
\begin{equation}
    H_{\rm hopping}^{\hat x+,\, \0+} = H_{\rm hopping}^{\hat x-,\, \0-} =\, {\rm diag}\{ t_p, t_{d1}, t_{d2} \},
\end{equation}
where the superscript $\RR x\pm$ represents the 3 basis states in the $j_z = \pm3/2$ sector at site $\RR$. Note that the two sectors are equivalent via time-reversal symmetry.
For inter-sector hopping, the parity and spin selection rules allow for only one independent nonzero hopping amplitude: $\braoperket{\hat x, d_{-2}\up}{H}{\0, p_{+1}\up} = t_{pd}$, giving
\begin{equation}
    H_{\rm hopping}^{\hat x-,\, \0+} = \begin{pmatrix}
        0 & 0 & t_{pd} \\ 0 & 0 & 0 \\ t_{pd} & 0 & 0
    \end{pmatrix}
\end{equation}
(note that the two nonzero elements have the same sign in the standard sign convention $p_{\pm1} \sim \mp(x\pm iy)$ and $d_{\pm2} \sim +(x\pm iy)^2$).
From Eqs. (\ref{eqS:d,+}), (\ref{eqS:d,-}) and the Slater-Koster relations \cite{slater1954simplified}, the regular and effective spin-flipping hopping amplitudes are respectively
\begin{equation}
\begin{aligned}
    t_{1, \tilde d} = \Braoperket{j_z = +\frac{3}{2}}{H_{\rm hopping}^{\hat x+,\, \0+}}{j_z = +\frac{3}{2}} &= t_{d1}\cos^2\theta + t_{d2}\sin^2\theta \\\addlinespace
    &= \frac{t_{dd\pi} + t_{dd\delta}}{2} \cos^2\theta + \frac{\frac{3}{4}t_{dd\sigma} + t_{dd\pi} + \frac{1}{4}t_{dd\delta}}{2} \sin^2\theta
\end{aligned}
\end{equation}
(here we have ignored the $\eta^2$ term) and
\begin{equation}
\begin{aligned}
    t_{1, \tilde d}^R = \Braoperket{j_z = -\frac{3}{2}}{H_{\rm hopping}^{\hat x-,\, \0+}}{j_z = +\frac{3}{2}} &= 2\eta t_{pd} \cos\theta\sin\theta  \\\addlinespace
    &= \lp( \frac{\sqrt{3}}{2}t_{pd\sigma} - t_{pd\pi} \rp) \eta\cos\theta \sin\theta.
\end{aligned}
\end{equation}
We see that in order to suppress $t_{1, \tilde d}$, we need $t_{d1}$ and $t_{d2}$ to have opposite signs, which is achievable within the regime of normal signs and hierarchy of Slater-Koster hopping amplitudes: $-t_{dd\sigma} > t_{dd\pi} > -t_{dd\delta} > 0$. Once $t_{d1}$ and $t_{d2}$ have opposite signs, $t_{1, \tilde d}^R$ can be suppressed by tuning $\theta$, which can in turn be engineered with $\lambda$, $\Delta_{pd}$ and $\Delta_d$ via Eq. (\ref{eqS:eta}). For instance, when $t_{dd\sigma} = -1$, $t_{dd\pi} = 0.5$ and $t_{dd\delta} = -0.2$, $t_{1,\tilde d}$ vanishes at $\theta = \pi/4$, which is satisfied when $3\lambda/2 + \eta\Delta_{pd} - \Delta_d = 0$.
We expect that relaxation of the assumptions $\Abs{\Delta_{pd}}, \Abs{\Delta_d}, \Abs{\lambda} \ll E_d - E_p$ does not affect the feasibility of engineering of hopping suppression.

\end{document}